\documentclass[prb,reprint,twocolumn,showpacs,superscriptaddress]{revtex4-1}
\usepackage{amsmath, amssymb, amsthm, braket}
\usepackage[breaklinks]{hyperref}
\usepackage{graphicx}
\usepackage{subfigure}
\usepackage{listings}

\newcommand{\hc}{{h.c.}}

\usepackage[usenames,dvipsnames]{color}
\definecolor{purple}{rgb}{0.5,0,0.5}

\begin{document}

\title{Stability of zero modes in parafermion chains}
\author{Adam S. Jermyn}
\affiliation{Department of Physics and Institute for Quantum Information and Matter, California Institute of Technology, Pasadena, California 91125, USA}
\author{Roger S. K. Mong}
\affiliation{Department of Physics and Institute for Quantum Information and Matter, California Institute of Technology, Pasadena, California 91125, USA}
\author{Jason Alicea}
\affiliation{Department of Physics and Institute for Quantum Information and Matter, California Institute of Technology, Pasadena, California 91125, USA}
\author{Paul Fendley}
\affiliation{Department of Physics, University of Virginia, Charlottesville, VA 22904-4714 USA}

\begin{abstract} 
One-dimensional topological phases can host localized zero-energy modes that enable high-fidelity storage and manipulation of quantum information. Majorana fermion chains support a classic example of such a phase, having zero modes that guarantee two-fold degeneracy in all eigenstates up to exponentially small finite-size corrections. Chains of `parafermions'---generalized Majorana fermions---also support localized zero modes, but, curiously, only under much more restricted circumstances.  We shed light on the enigmatic zero-mode stability in parafermion chains by analytically and numerically studying the spectrum and developing an intuitive physical picture in terms of domain-wall dynamics. Specifically, we show that even if the system resides in a gapped topological phase with an exponentially accurate ground-state degeneracy, higher-energy states can exhibit a splitting that scales as a \emph{power law} with system size---categorically ruling out exact localized zero modes. The transition to power-law behavior is described by critical behavior appearing exclusively within excited states.  
\end{abstract}

\maketitle

\section{Introduction}
Topological quantum computing represents a promising and conceptually elegant route to scalable quantum computation.\cite{kitaev,TQCreview}
Underlying this approach are topological phases of matter that harbor emergent particles known as non-Abelian anyons.  
Such particles exhibit two defining features:
	$(i)$ they generate a ground-state degeneracy that scales exponentially with the number of anyons present in the system,\cite{MooreRead:Nonabelion:1991,NayakWilczek:2nStatesQHPf96} and
	$(ii)$ braiding the anyons around one another non-commutatively rotates the system's quantum state within the ground-state manifold.\cite{MooreSeiberg:89, Witten:JonesPolynomial:1989}
The advantage that topological quantum computation offers over more traditional quantum computing schemes is that information is encoded and processed \emph{non-locally} in the braiding history of non-Abelian anyons.
Local environmental perturbations that ordinarily cause decoherence are thereby expected to be relatively benign.  

One conceptually simple realization of non-Abelian anyons are quasiparticles (or defects) that bind exponentially localized, topologically protected zero-energy modes.\cite{ReadGreen:p+ipFQHE:00, KitaevWireMajorana:01}
These modes are described by operators with appreciable weight on some finite length scale $\xi$ and that commute with the system's Hamiltonian up to exponentially small corrections $\sim e^{-R/\xi}$, where $R$ is the separation between adjacent anyons.
The localized character of the zero modes ensures well-defined braiding relations for the anyons that bind them, while the fact that they carry no energy guarantees the ground-state degeneracy necessary for non-Abelian statistics.
It is worth emphasizing, however, that zero modes so defined make an extremely strong statement about the system's spectrum: they imply an exponentially accurate degeneracy not just for ground states, but in fact for \emph{all} eigenstates.

\begin{figure}
\centering
\includegraphics[width=\columnwidth]{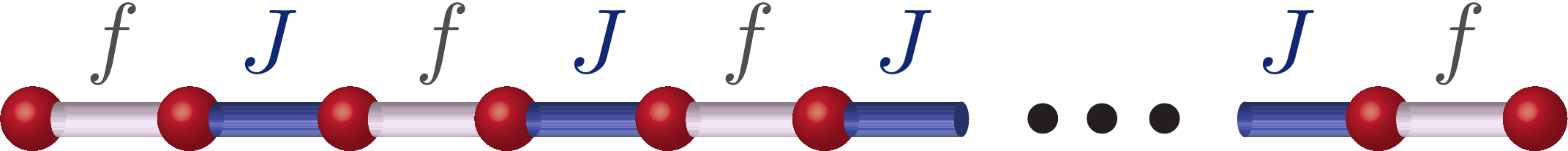}
\caption{Schematic illustration of either the Majorana or parafermion chain.  Adjacent sites couple with strength $f$ or $J$ as labeled above.  }
\label{fig:jfCouplingKitaev}
\end{figure}

The Kitaev chain \cite{KitaevWireMajorana:01} provides an illuminating example.
The Hamiltonian reads
\begin{align}
H=if\sum_x \gamma_{2x-1}\gamma_{2x}+iJ\sum_x \gamma_{2x}\gamma_{2x+1},
\end{align}
where $\gamma_x$ denotes a Hermitian Majorana fermion operator satisfying the commutation relation $\{\gamma_x,\gamma_{x'}\} = 2\delta_{x,x'}$.
As Fig.~\ref{fig:jfCouplingKitaev} illustrates, the couplings $f$ and $J$ favor competing Majorana dimerization patterns.  
In the special case with $f=0$, the outermost Majorana operators completely decouple from the rest of the system, commute with $H$, and thus form exact localized zero modes.  
Consequently, every eigenstate assumes at least two-fold degeneracy.  
Turning on finite $f$ preserves the zero modes---and with it the degeneracy in the spectrum---provided that $|f| < |J|$.\cite{KitaevWireMajorana:01}
Rather than localizing to one site the zero modes then simply decay exponentially into the bulk on the scale of the correlation length (which diverges at $|f| = |J|$).  
The survival of localized zero modes indicates that the chain resides in the same topologically nontrivial phase for any $|f| < |J|$.  
Throughout this phase the ends of the chain behave as non-Abelian anyons whose non-trivial exchange statistics can be meaningfully harvested in networks.\cite{AliceaBraiding,ClarkeBraiding,HalperinBraiding,BondersonBraiding}  
Braiding this type of anyon, however, enables only rather limited (i.e., non-universal) fault-tolerant quantum information processing.\cite{BravyiKitaev}

In the pursuit of non-Abelian anyons with greater utility for quantum computation, a variation of the Kitaev chain due to Fendley\cite{Fendley} has proven influential.  
The Hamiltonian for this `parafermion chain' (which Sec.~\ref{Model} discusses in depth) is given by
\begin{align}
  H=-f\sum_x\alpha_{2x-1}^\dagger\alpha_{2x}-J\sum_x \alpha_{2x}^\dagger\alpha_{2x+1}+\hc
  \label{Hpara}
\end{align}
Here $\alpha_x$ denotes a parafermion operator satisfying a $\mathbb{Z}_3$ generalization of the Majorana fermion algebra:
\begin{equation}
  \alpha_x^3 = 1,~~~~ \alpha_x^\dagger = \alpha^2_x,~~~~ \alpha_x \alpha_{x'>x} = e^{i2\pi /3} \alpha_{x'}\alpha_x.
  \label{alpha_properties}
\end{equation}
Figure~\ref{fig:jfCouplingKitaev} still illustrates the structure of the couplings, which we assume are real and non-negative throughout.  
With $f=0$ the outermost operators drop out from the Hamiltonian---precisely as in the Kitaev chain---and represent localized parafermion zero modes\footnote{To distinguish from parafermions in conformal field theory, which are related but distinct, these zero modes are sometimes referred to as `parafendleyons'.} that guarantee a three-fold degeneracy of every eigenstate.  

Similarities with the Majorana case, however, largely end here.  
Most strikingly, there is strong evidence that localized zero modes disappear entirely upon introducing \emph{arbitrarily small} $f$!\cite{Fendley}  
Such dramatic behavior defies intuition given that for $f< J$ the system resides in a gapped topological phase\cite{Fendley,Motruk,Bondesan} where one would naturally expect $f$ to yield only perturbative effects.
Stable localized zero modes were instead found only in a `chiral' deformation of the Hamiltonian wherein $J \rightarrow e^{i \phi} J$ with non-zero $\phi$ [more precisely, Fendley constructed localized zero modes when $f  \ll J |\sin (3\phi)|$].

Understanding the stark differences from the Kitaev chain and diagnosing implications for quantum information applications seem particularly pressing given the growing literature devoted to realizing parafermion zero modes 
(see, e.g., Refs.~\onlinecite{ChernInsulatorParafendleyons,ClarkeParafendleyons,LindnerParafendleyons,ChengParafendleyons,VaeziParafendleyons,BarkeshliParafendleyons1,BarkeshliParafendleyons2,BarkeshliClassification1,BarkeshliClassification2,Hastings,QuantumWiresParafendleyons,ParafendleyonLattice,Mong,Klinovaja1,Klinovaja2,BarkeshliDetection,Klinovaja3,ClarkeCircuits,ZhangKane,Orth,Li,Tsvelik}).
Despite all this work, the curious state of affairs regarding the stability of zero modes in parafermion chains has remained largely unexplored.  
The purpose of this paper therefore is to explain the generic absence of localized zero modes in Eq.~\eqref{Hpara} as well as their resurrection in the chiral case \cite{Fendley}.

Based on various complementary analytical and numerical methods, our work paints the following picture: In the topological phase with $f\neq 0$ the ground states, as expected, remain three-fold degenerate up to corrections that vanish exponentially as one approaches the thermodynamic limit.  Surprisingly, however, even the lowest-lying excited states that would otherwise be exactly degenerate at $f = 0$ exhibit a \emph{power-law} splitting with system size for $f \neq 0$, implying the destruction of localized zero modes.  We show that the onset of power-law splitting can be understood via domain-wall tunneling processes that simply have no analogue in the Kitaev chain.  We further demonstrate that chirally deforming the Hamiltonian frustrates these domain-wall tunneling events, eventually restoring exponential splitting of the states (at least in part of the spectrum) consistent with zero-mode revival.  

One noteworthy implication of our work is that the disappearance of localized zero modes should \emph{not} be conflated with a demise of non-Abelian anyons.  
On the contrary, throughout the topological phase exhibited by Eq.~\eqref{Hpara} the parafermion chain still allows one to demonstrate non-Abelian statistics since the all-important ground-state degeneracy persists modulo exponentially small corrections. 
For such cases it should be possible to define weaker zero mode operators that are localized and commute with a \emph{projected} Hamiltonian.\footnote{For an explicit construction see A.~Alexandradinata, C.~Fang,  N.~Regnault,  and B.~A.~Bernevig, unpublished.}  
As another interesting corollary, we show that the transition between power-law and exponential splitting noted above can be associated with chirality-tuned critical behavior in the \emph{excited states}, even though the ground state sector remains regular.  

For completeness we note that parafermion braiding, while carrying some advantages over the Majorana case, remains non-universal.\cite{ClarkeParafendleyons,LindnerParafendleyons,Hastings}
One can, nevertheless, leverage parafermionic systems to generate new two-dimensional phases that do permit universal topological quantum computation.\cite{Mong}
Interestingly, similar physics can even appear in local bosonic two-dimensional systems.\cite{QiSlaveGenon}.

To flesh out the above results, we begin in Sec.~\ref{Model} by describing basic properties of the parafermion chain model---in particular explaining a non-local mapping to `spins' of the three-state Potts model---and introduce the criteria used for evaluating the existence of zero modes.  
Sections \ref{PerturbativeRegime} and \ref{sec:nonperturb} then explore the ground states and excited states of the Hamiltonian using perturbation theory, exact diagonalization of a truncated Hilbert space model, and density matrix renormalization group (DMRG) simulations.  
Finally, we conclude in Sec.~\ref{Conclusions} by highlighting additional implications and extensions of this study.

\section{The Model and Zero-Mode Criterion}
\label{Model}

The most general parafermion chain Hamiltonian studied in this paper is given by
\begin{align}
	H=-f\sum_{x = 1}^{L}\alpha_{2x-1}^\dagger\alpha_{2x}-Je^{i\phi}\sum_{x = 1}^{L-1} \alpha_{2x}^\dagger\alpha_{2x+1}+\hc,
	\label{eq:Halpha}
\end{align}
where again $\alpha_x$ satisfies the properties in Eq.~\eqref{alpha_properties}.
Note that in total the system consists of $2L$ parafermion sites (to define a sensible Hilbert space this number is necessarily even).  
Without loss of generality we will restrict the chiral phase $\phi$ appearing in the second term to the range $\phi \in [0,\pi/3]$, since symmetry relates Hamiltonians with $\phi \rightarrow \phi + 2\pi/3$ and $\phi \rightarrow -\phi$ to Eq.~\eqref{eq:Halpha}.

As noted above, in the limit $f = 0$ the existence of zero modes at the ends of the chain is obvious since then $[H, \alpha_1] = [H,\alpha_{2L}] = 0$ for any $J, \phi$.  
To appreciate the implications of these zero modes it is useful to define a `triality' operator 
\begin{equation}
  \hat{Q} = \prod_{x = 1}^L \alpha_{2x-1}^\dagger\alpha_{2x}
\end{equation}  
akin to the total fermion parity in the Kitaev chain.  
Since $\hat{Q}^3=1$,  $\hat{Q}$ admits eigenvalues $1$, $\omega$, or $\omega^2$,
where
\begin{align}
	\omega = e^{i2\pi/3}.
\end{align}
For any choice of couplings $\hat{Q}$ commutes with the Hamiltonian. 
Crucially, however, the zero-mode operators $\alpha_{1}$ and $\alpha_{2L}$ do \emph{not} commute with $\hat Q$---they cycle the triality by $\omega$.
It follows that the entire spectrum can be grouped into triplets of energy eigenstates with trialities $Q = 1$, $\omega$, and $\omega^2$ that are exactly degenerate at $f = 0$.  
The spectrum of the Kitaev chain in the analogous limit consists of degenerate doublets with opposite fermion parity.

Our goal is to explore the fate of these localized zero-mode operators in the generic situation with $f \neq 0$.  
We now define the precise criteria used in evaluating whether or not exact edge zero modes exist.
Finite-size effects at nonzero $f$ generically split the exact degeneracy between different triality states except with fine-tuning.  
Let $E_{a,Q}$ denote the system's energies, where $Q$ labels the triality and $a = 1,2,3,\ldots$ indexes the levels such that $E_{1,Q} \leq E_{2,Q} \leq E_{3,Q} \cdots$.  
The existence of exponentially localized zero modes implies that $|E_{a,Q}-E_{a,Q'}|=\mathcal{O}\left(e^{-L/\xi}\right)$ holds for every $a,Q,Q'$ with some length scale $\xi>0$.
An exceedingly useful corollary is that the existence of such modes may be categorically ruled out in large swaths of parameter space merely by demonstrating sub-exponential (e.g., power-law) splitting of a single triplet of energy levels $E_{a,Q = 1,\omega,\omega^2}$.
Demonstrating the presence of exact zero modes, by contrast, poses a much more difficult problem, as doing so requires proving a global property of all energy levels.  
In this paper we will content ourselves with identifying regimes where zero modes are definitely absent.

One can obtain a great deal of intuition by exploiting an exact mapping between the parafermion chain Hamiltonian in Eq.~\eqref{eq:Halpha} and the chiral three-state Potts model. This mapping is analogous to that between the Kitaev chain and the transverse-field Ising model, and accordingly is implemented with a variation of the Jordan-Wigner transformation introduced by Fradkin and Kadanoff\cite{FradkinKadanoff}.  In particular, writing
\begin{align}
	\alpha_{2x-1} &= \sigma_x\prod_{j<x}\tau_j,
&	\alpha_{2x} &= \sigma_x\prod_{j\leq x}\tau_j.
  \label{PottsMapping}
\end{align}
decomposes the parafermions via strings of Potts model `spin' operators that satisfy
\begin{align}
	\sigma_x^3&=\tau_x^3=1,  &\sigma_x\tau_x&= \omega\tau_x\sigma_x.  
	\label{PottsAlgebra}
\end{align}
All other commutators amongst $\sigma_x$ and $\tau_x$ vanish, a straightforward result of Eq.~\eqref{alpha_properties}.
Under this non-local change of variables the Hamiltonian becomes\footnote{In writing the Potts representation of the parafermion chain, we dropped a factor of $e^{-i2\pi/3}$ in the $J$ term; this factor can always be absorbed into the chiral phase $\phi$.}
\begin{align}
	H=-Je^{i \phi}\sum_{x = 1}^{L-1} \sigma_x^\dagger\sigma_{x+1}-f\sum_{x = 1}^L\tau_x+\hc,
	\label{Hpotts}
\end{align}
which defines a local bosonic model whose states are much easier to analyze than those in the parafermion chain.
We stress, however, that both models are equivalent and exhibit precisely the same spectra.
Equation \eqref{Hpotts} can be obtained from the anisotropic limit of a well-known classical lattice model. With $\phi=0$ and $J > 0$ this is the ferromagnetic three-state Potts model.  When $\phi\ne 0$, this is typically called the chiral clock or sometimes the chiral Potts model.

Using a basis of $\sigma$ eigenstates, denoted $|s = 0,1,2\rangle$, the Potts operators on a given site can be represented as
\begin{align}
	\sigma=\begin{pmatrix}
		1 & 0 & 0 \\
		0 & \omega & 0 \\
		0 & 0 & \omega^2 \end{pmatrix} ,\quad
	\tau=\begin{pmatrix}
		0 & 1 & 0 \\
		0 & 0 & 1 \\
		1 & 0 & 0 \end{pmatrix} .
\end{align}
We then have
\begin{align}
	\sigma\ket{s} = \omega^s\ket{s} ,
	\quad	\tau\ket{s} = \ket{s+1\bmod3} ,
\end{align}
from which it follows that $\tau$ cycles the `spin' measured by $\sigma$.  
Notice that $\sigma$ and $ \tau$ represent a straightforward generalization of anticommuting Pauli matrices in the Ising model.
The triality defined earlier is simply the generator of a global $\mathbb{Z}_3$ symmetry exhibited by the Potts Hamiltonian, winding every spin:
\begin{align}
	\hat{Q} = \prod_x \tau_x \ .
	\label{Qpotts}
\end{align}

Hereafter we find it much more illuminating to work in the Potts model representation.  
Thus it is worth translating into Potts language the consequences of localized zero modes in the parafermion chain.  
For simplicity, consider $\phi = 0$ and suppose for the moment that $f = 0$---where parafermion zero modes definitely exist.  
In this case Eq.~\eqref{Hpotts} reduces to the ferromagnetic Potts model with a vanishing `transverse field'.
The Hamiltonian has three broken-symmetry ground states that transform into one another under the action of $\hat{Q}$ in Eq.~\eqref{Qpotts}; we label these by $|A\rangle = |00\cdots0\rangle$, $|B\rangle = |11\cdots1\rangle$, and $|C\rangle = |22\cdots2\rangle$.  
One can of course also define a basis of Schr\"odinger-cat-like ground states with definite triality via
\begin{align}\begin{split}
	|Q = 1\rangle &= |A\rangle + |B\rangle + |C\rangle ,
\\	|Q = \omega\rangle &= |A\rangle + \omega |B\rangle + \omega^2|C\rangle ,
\\	|Q = \omega^2\rangle &= |A\rangle + \omega^2|B\rangle + \omega|C\rangle .
   \label{Qgdstates}
\end{split}\end{align}
The zero-mode operators $\alpha_1$ and $\alpha_{2L}$ cycle the system amongst the degenerate triplet defined in Eqs.~\eqref{Qgdstates} [to see this recall Eq.~\eqref{PottsMapping}].  
Similar conclusions hold for the excited states, which one can fruitfully view in terms of domain walls (i.e., kinks and anti-kinks) separating different ferromagnetically ordered regions.
All such excited states may also be grouped into degenerate triplets of
triality eigenstates that transform into one another under the action of the localized parafermion zero-mode operators.  Analogous results apply to the chiral case with $\phi \neq 0$ despite the fact that chirality can nontrivially rearrange the spectrum (see the next section).

Restoring non-zero $f$ lifts the degeneracy among the ground-state and excited triplets via a transparent physical mechanism.  
Namely, the $f$ term in Eq.~\eqref{Hpotts} creates mobile domain walls that can tunnel the system between states related by a global $\mathbb{Z}_3$ transformation.  
Repeated action of the $f$ term can, for instance, send $|A\rangle \rightarrow |B\rangle$, $|B\rangle\rightarrow |C\rangle$, and $|C\rangle \rightarrow |A\rangle$, thereby splitting the three-fold ground-state degeneracy in the ferromagnetic case.  The question we address in the following sections is thus: {\em how} do the degeneracies split?

\section{Perturbative Regime}
\label{PerturbativeRegime}

With the insights of the previous section we now distill our earlier criterion for the existence of localized zero modes to the following question: 
Do domain-wall hopping processes preserve the degeneracy amongst \emph{all} triplets of energy eigenstates, up to corrections that decay to zero exponentially with system size?
If the answer is `no' then localized zero modes in the parafermion chain are ruled out.  
In this section we address this question by studying both the splitting of ground states and of low-lying excited states in the limit $f/J \ll 1$.  Section~\ref{sec:nonperturb} then explores the complementary regime where $f/J$ is of order one.

\begin{figure}
	\centering
	\includegraphics[width=67mm]{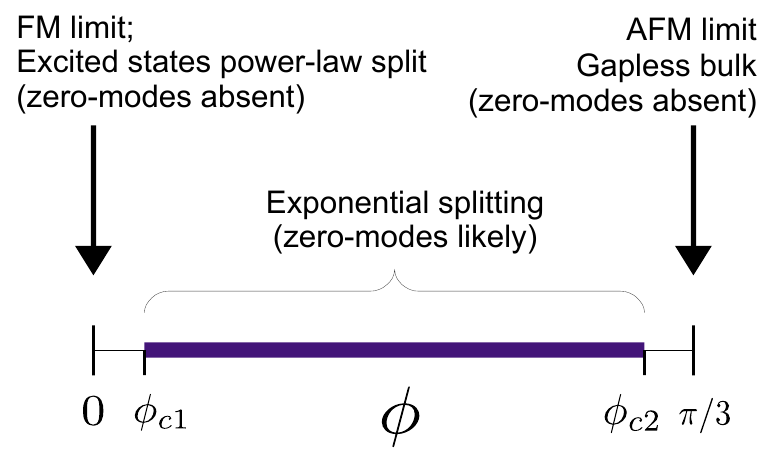}
	\caption{
		Summary of results for the perturbative regime $f\ll J$.
		The two limits $\phi = 0$ and $\phi = \pi/3$ respectively correspond to the ferromagnetic and antiferromagnetic points of the chiral Potts model [Eq.~\eqref{Hpotts}].  Note that the critical values $\phi_{c1}$ and $\phi_{c2}$ depend on $f/J$.  
	}
	\label{fig:smallFphases}
\end{figure}

As described above the entire spectrum consists of triplets of exactly degenerate states when $f = 0$.
With $f/J \ll 1$, degenerate perturbation theory allows one to quantify the effect of non-zero $f$ on these triplets.
In what follows we discuss the ferromagnetic model with $\phi = 0$ from this perturbative perspective, then attack the chiral case with $\phi \neq 0$, and finally address the antiferromagnetic limit $\phi = \pi/3$.  
In both the ferromagnetic and antiferromagnetic limits we will show that localized zero modes certainly do not exist even for arbitrarily small but finite $f$, consistent with Ref.~\onlinecite{Fendley}.  
Rigorous conclusions are harder to obtain for the chiral case, though in our analysis we definitively rule out zero modes over ranges of $\phi$ and $f$ in the vicinity of $\phi = 0$ and $\pi/3$; 
we argue that outside of these regimes chirality restores zero modes in the perturbative regime.  
Figure \ref{fig:smallFphases} summarizes our results for this section.

\subsection{Ferromagnetic limit, \texorpdfstring{$\phi = 0$}{phi = 0}}
\label{FMpotts}

\begin{figure*}
	\centering
	\includegraphics[width=0.85\textwidth]{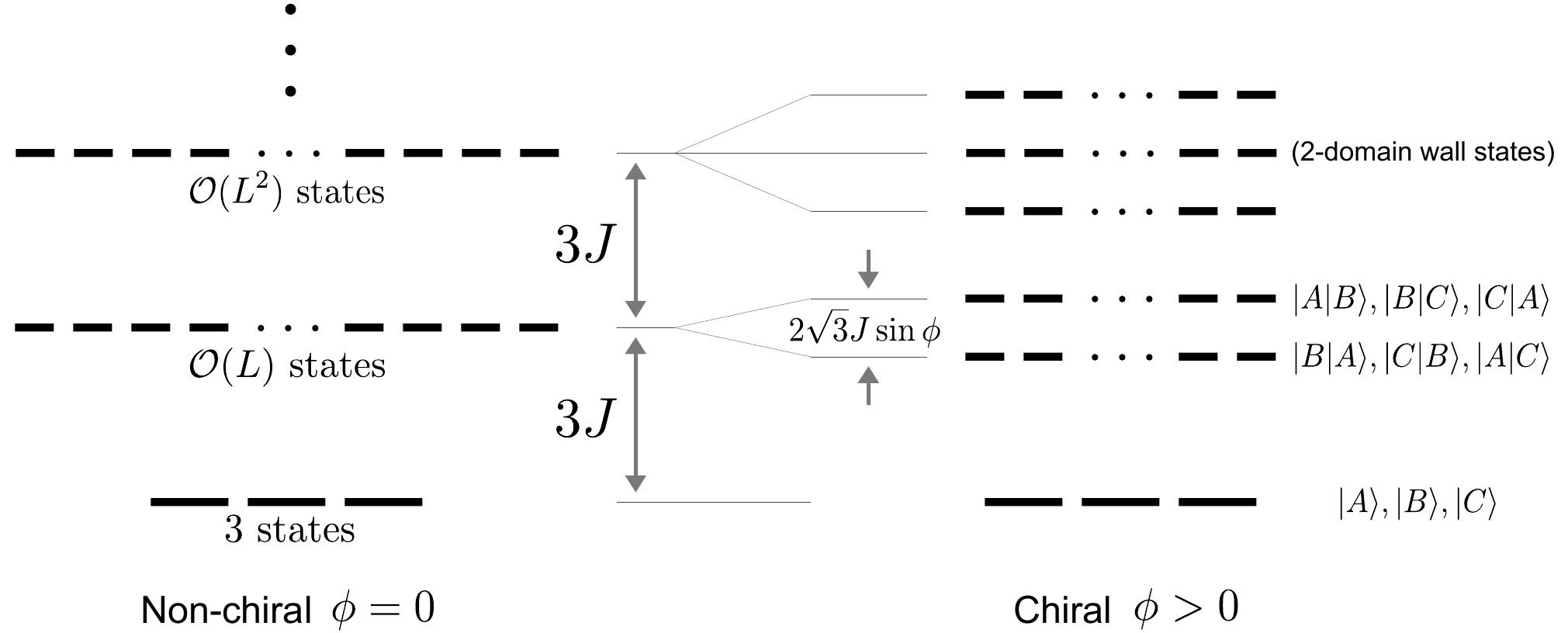}
	\caption{
		Schematic illustration of the $f = 0 $ level spectrum in the ferromagnetic limit (left half) and in the chiral regime for small $\phi$ (right half).
	        The ferromagnetic limit supports a series of highly degenerate excited states obtained by inserting domain walls into one of the three ground states.  
	        Introducing non-zero chirality preserves the three ground states but partially splits the excited-state degeneracy as shown on the right.  	       	       
	}
	\label{fig:spectrum}
\end{figure*}

Figure \ref{fig:spectrum} (left side) illustrates the $f = 0$ spectrum for the ferromagnetic case.  
Let us now discuss these levels in greater detail.  
There are three ground states $|A\rangle, |B\rangle$, and $|C\rangle$ corresponding to all spins uniformly polarized along one of three possible directions.
As in the Ising model every excited state can be viewed in terms of domain walls between differently polarized regions, although here six flavors exist rather than two because of the larger ground-state degeneracy.  The energy cost of a domain wall comes from the single link connecting different values of the spins, giving
\begin{equation}
  E_{\rm wall} \Big|_{f=0} = 2J\left[1-\cos\left(2\pi/3\right)\right] = 3J.
\end{equation}
The lowest excited levels of Fig.~\ref{fig:spectrum} are single-domain-wall states. The six flavors are denoted as $|A|B\rangle,$ $|B|C\rangle$, $|C|A\rangle$, $|B|A\rangle, |C|B\rangle$, and $|A|C\rangle$, where for example $|A|B\rangle$ indicates the presence of a domain wall where spins on the left are in ground state $|A\rangle$ and on the right are in $|B\rangle$. Each flavor can sit at any of $L-1$ positions, so a total of $6(L-1)$ such states exist---all exactly degenerate at $f = 0$.  
There exists a larger set of $\mathcal{O}(L^2)$ degenerate two-domain-wall states corresponding to $|A|B|C\rangle, |A|B|A\rangle, |C|B|A\rangle$, etc., and so on up the spectrum.
Crucially, the three ground states and each set of degenerate domain-wall states form subspaces that are energetically well-separated from one another by a gap $3J$.  

Including infinitesimal $f$ enables domain-wall creation, annihilation, and hopping. The splitting of the levels within a given subspace can then be computed using standard degenerate perturbation theory methods. We are particularly interested in processes that take one $f = 0$ eigenstate to another related by a global $\mathbb{Z}_3$ transformation, e.g., $|A\rangle \rightarrow |B\rangle$ or $|A|B\rangle \rightarrow |B|C\rangle$.  
These are precisely the processes that produce finite-size splitting of the degeneracy encoded by any localized zero modes.

\begin{figure}
\centering
\subfigure[]{
\includegraphics[width=0.23\textwidth]{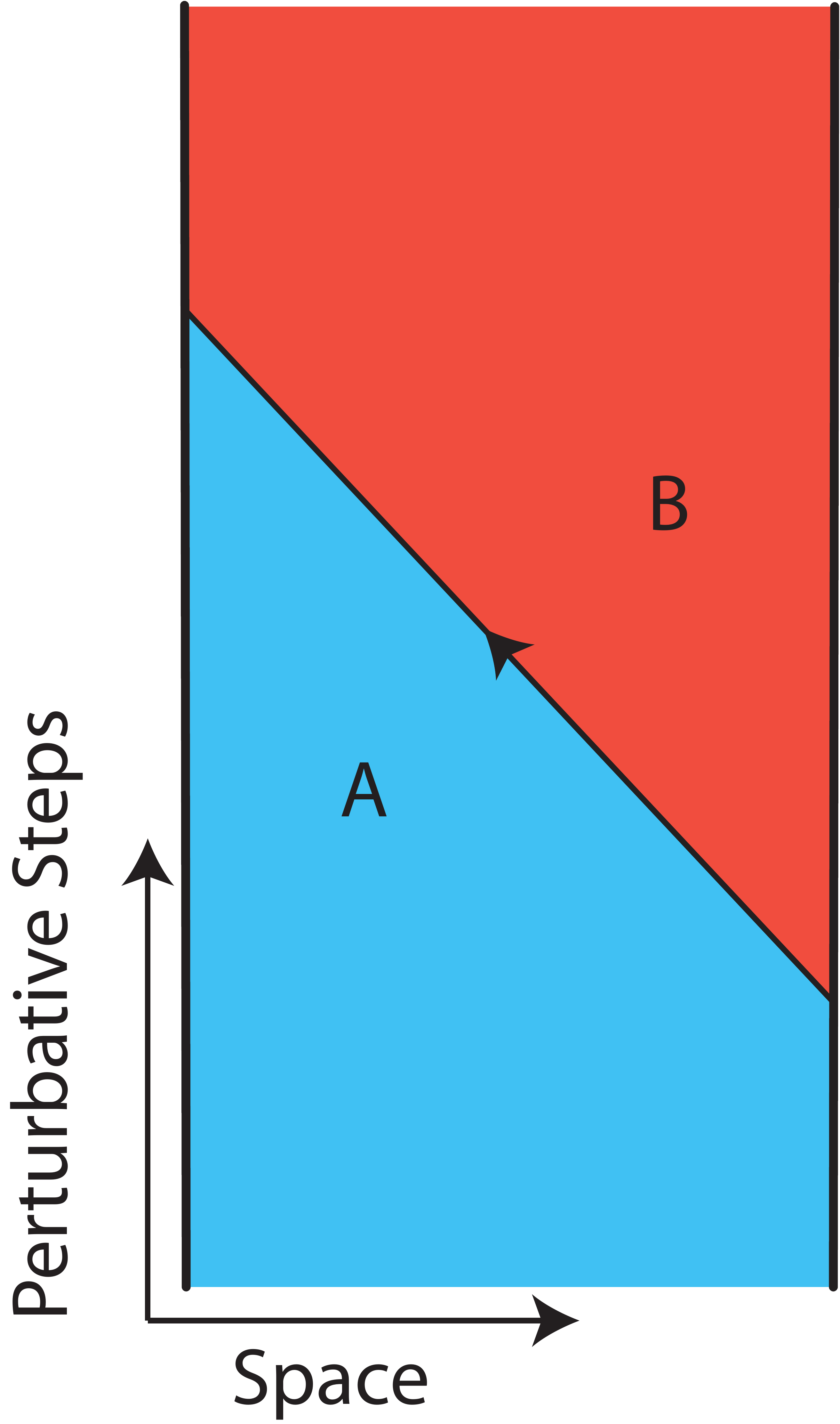}
\label{fig:diag2}}\qquad
\subfigure[\, ]{
\includegraphics[width=0.18\textwidth]{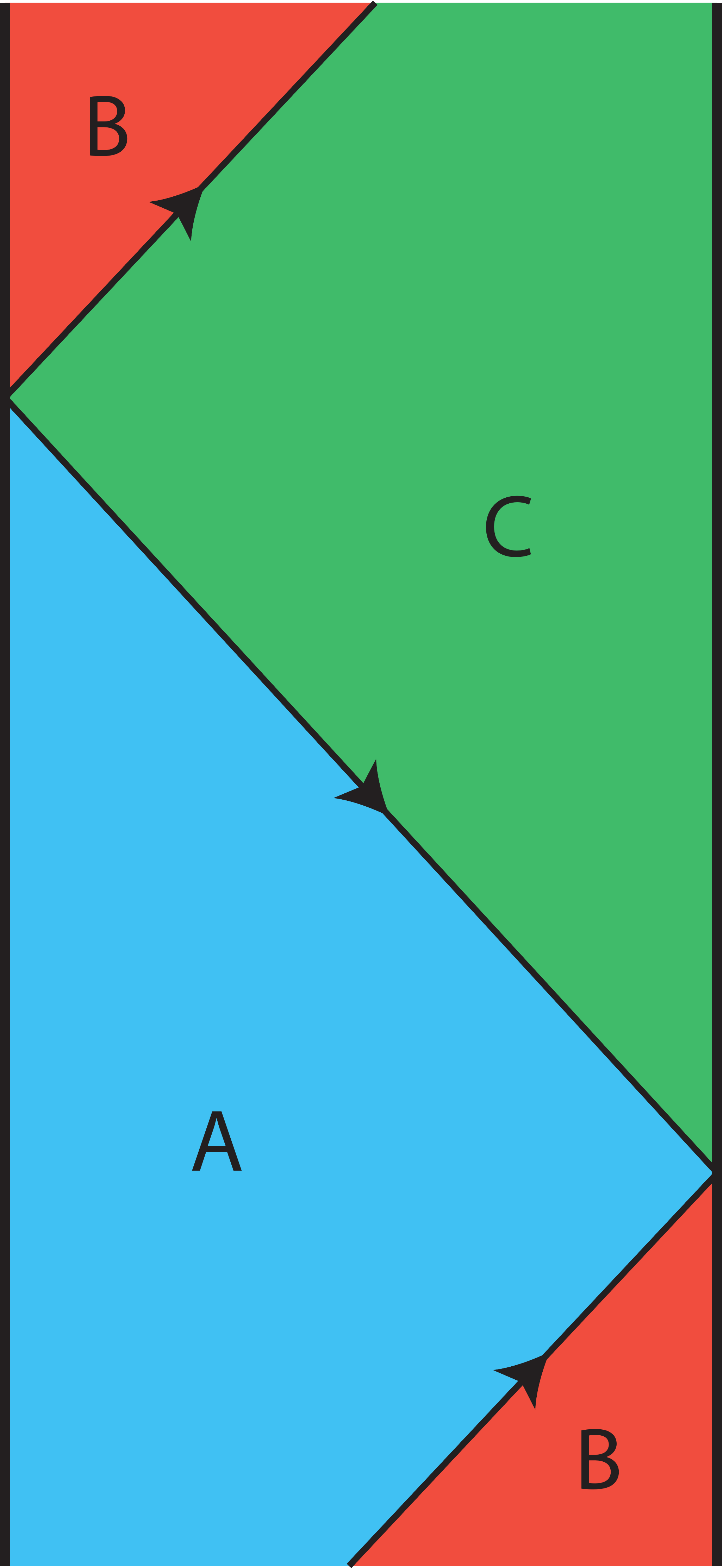}
\label{fig:oneWallDiag1}}
\caption{Examples of processes that connect $f = 0$ eigenstates related by a global $\mathbb{Z}_3$ transformation.  
Such processes are particularly important as they generate splitting among triplets of states that are degenerate at $f = 0$, reflecting the exact zero modes present in that limit.  
In (a) the ground state $|A\rangle$ converts into $|B\rangle$ via tunneling of a domain wall from the right to the left end of the system.  
This requires spending a macroscopic number of steps in an excited state, leading to exponential splitting.  
In (b) an excited state transitions from $|A|B\rangle\rightarrow |A|C\rangle \rightarrow |B|C\rangle$ while remaining entirely in the single-wall subspace.  
For the ferromagnetic case $\phi = 0$ power-law splitting among the excited triplets thus arises, precluding exact zero modes for arbitrarily small but finite $f$.  
With any chirality $\phi$ between $0$ and $\pi/3$, there exists $f$ sufficiently small that exponential splitting reappears, since the transitions to and from $|A|C\rangle$ happen non-resonantly for all excited states.  
At larger $f$ power-law splitting again emerges from such processes since there is a finite energy window in which the different domain-wall bands overlap.  
See Appendix \ref{Diagrams} for the meaning of the arrows.
}
\end{figure}

The splitting of the ground-state energies due to non-zero $f$ follows from standard arguments.  These imply that mixing between the three ground states $|A\rangle, |B\rangle, |C\rangle$ is suppressed \emph{exponentially} in system size. Namely, the `cheapest' way to evolve from, say, $|A\rangle$ to $|B\rangle$ is to $(i)$ create an $|A|B\rangle$ domain wall at one of the end chain---leaving the system in an excited state, $(ii)$ tunnel the wall over to the opposite end, and $(iii)$ annihilate the domain wall to re-enter the ground-state manifold.  
One can visualize the process graphically via Fig.~\ref{fig:diag2}.  
Here the horizontal axis denotes the position along the chain, the vertical axis represents the perturbation step (which can also be interpreted as time), and the diagonal line indicates the domain-wall trajectory.  
Such a process necessitates exiting the ground-state subspace for a macroscopic number of steps, resulting in a ground-state splitting 
\begin{equation}
  \Delta E_{\rm g.s.} \sim f\left(\frac{f}{3J}\right)^{L-1}.
  \label{Egs}
\end{equation} 
In short, an energy barrier prevents efficient mixing of these states through local spin flips.  
Thus the \emph{ground-state} degeneracy encoded by the exact zero modes of the $f = 0$ limit survives the introduction of non-zero $f$, up to exponentially small corrections in system size.  

Qualitatively different behavior arises in the single-domain-wall sector.  
In fact here the energy barrier that suppresses ground-state mixing disappears altogether so that the dominant contribution to the splitting that we seek arises already at \emph{first order} in degenerate perturbation theory.  
To see this consider the process illustrated in Fig.~\ref{fig:oneWallDiag1} which takes $|A|B\rangle \rightarrow |B|C\rangle$ without leaving the single-wall subspace.  
This process proceeds by first hopping the $|A|B\rangle$ domain wall all the way to the right end of the chain.  
If we were dealing with an Ising model, then the only way to remain in the original subspace would be to subsequently tunnel the domain wall leftward---which fails to accomplish the domain-wall change required.
The Potts chain, however, offers an alternative: the rightmost spin can wind, converting $|A|B\rangle$ to $|A|C\rangle$.  This newly formed $|A|C\rangle$ domain wall can then tunnel all the way to the chain's left end where it can similarly convert to $|B|C\rangle$.  Moving the domain wall back to its original location completes the process.

The splitting amongst triplets of single-domain-wall states resulting from such processes certainly depends sub-exponentially on system size.  Indeed, the energy denominators responsible for the exponential dependence in Eq.~\eqref{Egs} are entirely absent since the system never leaves the degenerate single-wall subspace (again, all the action takes place at first order in perturbation theory).
We can explore the splitting more quantitatively by examining the Hamiltonian projected onto the single-domain-wall subspace, which resembles a tight-binding model for six flavors of particles reflecting the domain-wall types; for details see Appendix \ref{ProjectedH}.
If one were to ignore the fact that these particles are domain walls and give them periodic boundary conditions, then the energy would be
\begin{equation}
  E(k) =  3J-2f\cos k,
  \label{Ek}
\end{equation}
where $k \in 2\pi \mathbb{Z}/L$ labels the momentum.
These energy levels are trivially six-fold degenerate since the domain-wall flavor is conserved with periodic boundary conditions.  
With open boundaries, however, the system's edges mediate backscattering processes that \emph{do} convert domain-wall flavors into one another as in Fig.~\ref{fig:diag2}.
Since the backscattering happens within a continuum of extended levels, power-law splitting generically arises for all triplets of single-wall states even with arbitrarily small non-zero $f$.

We can compute analytically the power-law splitting arising with open boundaries in the single-domain-wall projection. 
The key observation is that there is only one way for domain walls to change flavor at each end, e.g., $|A|B\rangle$ to $|A|C\rangle$ at the right boundary. 
As a consequence, we can `unfold' the system into a \emph{periodic} chain of size $6(L-1)$ with a single flavor. 
In this unfolded picture each flavor of the original model corresponds to a region of size $L-1$ so that a domain wall bouncing from an edge is replaced by a particle moving into a different region.  
Cycling through all six flavors corresponds to going around the periodic chain.  

\begin{figure}
	\centering
	\includegraphics[width=\columnwidth]{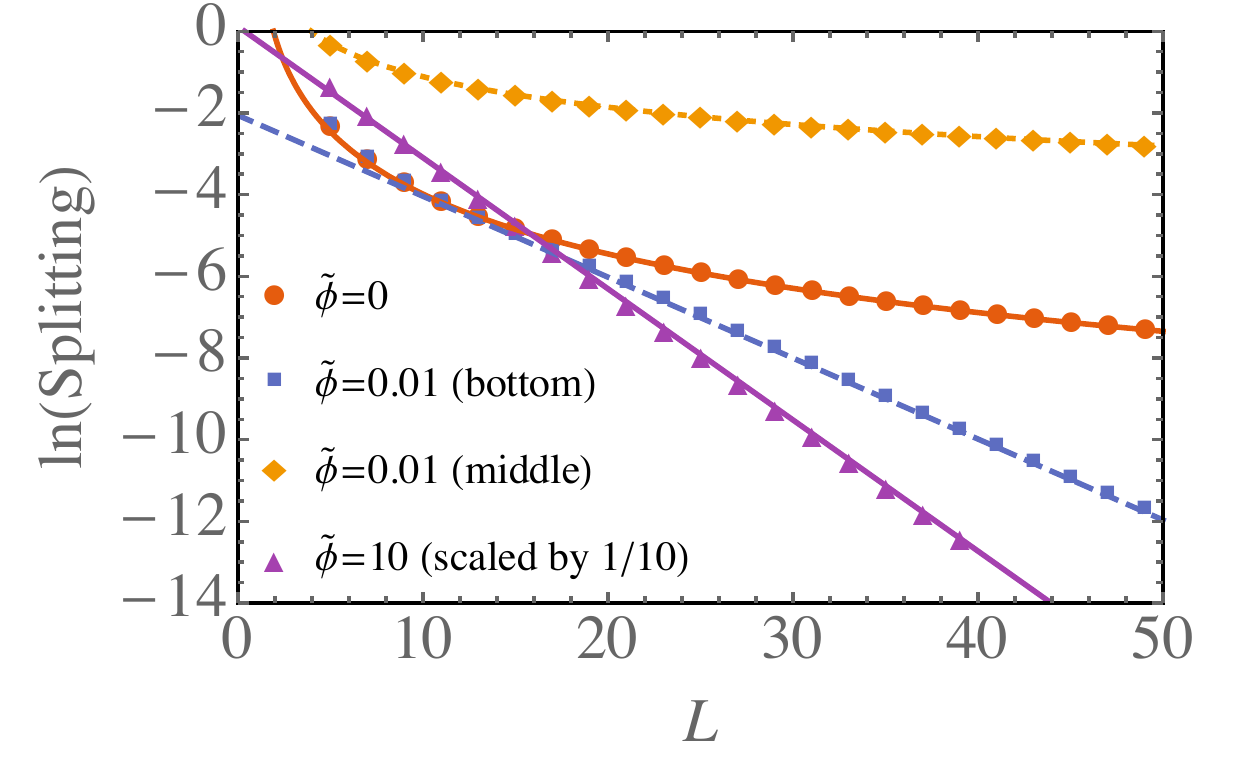}
	\caption{
		Exact diagonalization results for the Hamiltonian projected into the single-domain-wall subspace with various values of $\tilde{\phi} = \sqrt{3} J\sin\phi/f$.  
	The vertical axis denotes the natural log of the splitting for a given triplet, defined as in Eq.~\eqref{Splitting}.
	All data correspond to the lowest excited triplet except for the diamonds, which represent a triplet at the center of the lowest single-domain-wall band.
	The $\tilde{\phi}=10$ data has been rescaled by a factor of $1/10$ for visual clarity.
	With $\tilde \phi = 0$, power-law splitting arises that fits extremely well to a $1/L^2$ curve (solid line through circles).  
		Similarly, the $\tilde\phi = 0.01$ data corresponding to the band middle exhibit power-law splitting consistent with a $1/L$ scaling (dotted line through diamonds).
		Other finite-$\tilde\phi$ data sets exhibit exponential decay as shown by the fits (dashed and dotted lines through squares and triangles).
}
	\label{fig:oneWallED}
\end{figure}

We can then obtain the spectrum using Fourier analysis; details of the calculation appear in Appendix \ref{ProjectedH}. 
The solution exhibits an interesting form that can be anticipated from previously known results for the scaling limit of the three-state Potts model:  
the amplitude for changing flavor when a kink scatters off the boundary is of magnitude 1, while the amplitude for not changing (i.e., just bouncing back as is) vanishes \cite{Chim}.  
Indeed we find that the eigenstates take the form of simple right- or left-moving plane waves that propagate unreflected around the unfolded periodic chain.  
More quantitatively, the results are as follows.
Labeling the energies of a single domain-wall triplet as $E_{a,Q = 1,\omega,\omega^2}$ for integer $a=1\dots L-1$, we define
\begin{equation}
  {\rm Splitting}[a] \equiv \sqrt{\sum_{Q' < Q} (E_{a,Q}-E_{a,Q'})^2}.
  \label{Splitting}
\end{equation}
In the large-$L$ limit this becomes
\begin{equation}
{\rm Splitting}[a] \approx \frac{2\sqrt{2} \pi f}{3L}\sin\left(\frac{\pi a}{L}\right)\ .
\label{eq:splittEq}
\end{equation}
Near the band edges a power-law of the form $1/L^2$ emerges; elsewhere a $1/L$ scaling takes over. 
We verified the splittings extracted above by numerically simulating the projected Hamiltonian with open ends; very large system sizes can be readily simulated since the truncated Hilbert-space dimension scales only linearly with $L$.  
Figure~\ref{fig:oneWallED} (red circles) illustrates our simulation results.
The vertical axis denotes the natural log of the splitting of the three lowest-energy domain-wall eigenstates while the horizontal axis represents $L$.  
Clear $1/L^2$ splitting indeed appears as shown by the solid line.

The power-law behavior captured here immediately precludes the existence of exponentially localized zero modes in the ferromagnetic limit for \emph{any} small but finite $f/J$.  
We stress that this result is quite robust as processes that involve neglected subspaces are higher-order in $f$ and in the perturbative regime cannot remove the power-law splitting arising at first order.

\subsection{Chiral case, \texorpdfstring{$0 < \phi < \pi/3$}{0 < phi < pi/3}}
\label{Chiralpotts}

Having ruled out exact localized zero modes in the ferromagnetic case, we turn now to the chiral regime where $\phi$ lies between 0 and $\pi/3$ non-inclusive.  
It is instructive to first discuss the $f = 0$ spectrum illustrated schematically on the right side of Fig.~\ref{fig:spectrum}.  
For any $\phi$ in this range $|A\rangle, |B\rangle, |C\rangle$ remain the unique exact ground states. Chirality does, however, alter the excited states in an important way.
More precisely, the $|A|B\rangle$, $|B|C\rangle$, and $|C|A\rangle$ domain walls now carry energy $E^+_{\rm wall}(\phi)$ that differs from the energy $E^-_{\rm wall}(\phi)$ carried by their mirror counterparts $|B|A\rangle$, $|C|B\rangle$, and $|A|C\rangle$.
Explicitly,
\begin{equation}
  E^\pm_{\rm wall}(\phi) = 2J\left[\cos\phi-\cos\left(2\pi/3\pm\phi\right)\right].
  \label{Echiral}
\end{equation}
The single-domain-wall sector therefore splits into two degenerate subspaces, the two-wall sector splits into three ($|A|B|C\rangle$, $|C|B|A\rangle$ and $|A|B|A\rangle$ all yield different energy), etc.
Note that Fig.~\ref{fig:spectrum} illustrates the levels only for small chiral phases $\phi$; larger $\phi$ re-orders the excited states as we discuss below.  

Chirality imparts only minor quantitative effects on the ground-state splitting induced by non-zero $f$.  
Our arguments from the previous subsection indeed carry over straightforwardly since a finite gap $\Delta(\phi) = E^-_{\rm wall}(\phi)$ to the first excited state persists whenever $0<\phi<\pi/3$.
Exponentially small splitting is thus again guaranteed over a range of $f$, the primary difference being that the denominator $3J$ in Eq.~\eqref{Egs} should be replaced by the gap $\Delta(\phi)$. 

For the excited states, by contrast, chirality yields more dramatic consequences.  
This originates from the reduced degeneracy of the $f = 0$ domain-wall eigenstates relative to the ferromagnetic case, which tends to suppress the tunneling processes that previously led to power-law splitting.

It is simplest to examine small chiral phases $\phi$ where the system remains close to the ferromagnetic point; values of $\phi$ near the antiferromagnetic point $\pi/3$ will be discussed separately below.  
Following Sec.~\ref{FMpotts} we can capture the leading effects of $f$ on the low-lying excited states by projecting the Hamiltonian onto the single-domain-wall subspace.  
The effective Hamiltonian (see Appendix \ref{ProjectedH}) again resembles a tight-binding model for six flavors of particles, half of which now experience different on-site energies resulting in two types of excitation branches.  
With periodic boundary conditions one obtains band energies
\begin{equation}
  E_\pm(k,\phi) =  E^\pm_{\rm wall}(\phi)-2f\cos k
  \label{bands}
\end{equation}  
that exhibit three-fold degeneracy for each momentum $k$.  
The upper branch $E^+(k,\phi)$ represents the bands formed by $\{|A|B\rangle, |B|C\rangle, |C|A\rangle\}$ states while the lower branch similarly corresponds to $\{|B|A\rangle, |C|B\rangle, |A|C\rangle\}$.

\begin{figure}
\centering
\includegraphics[width= 7cm]{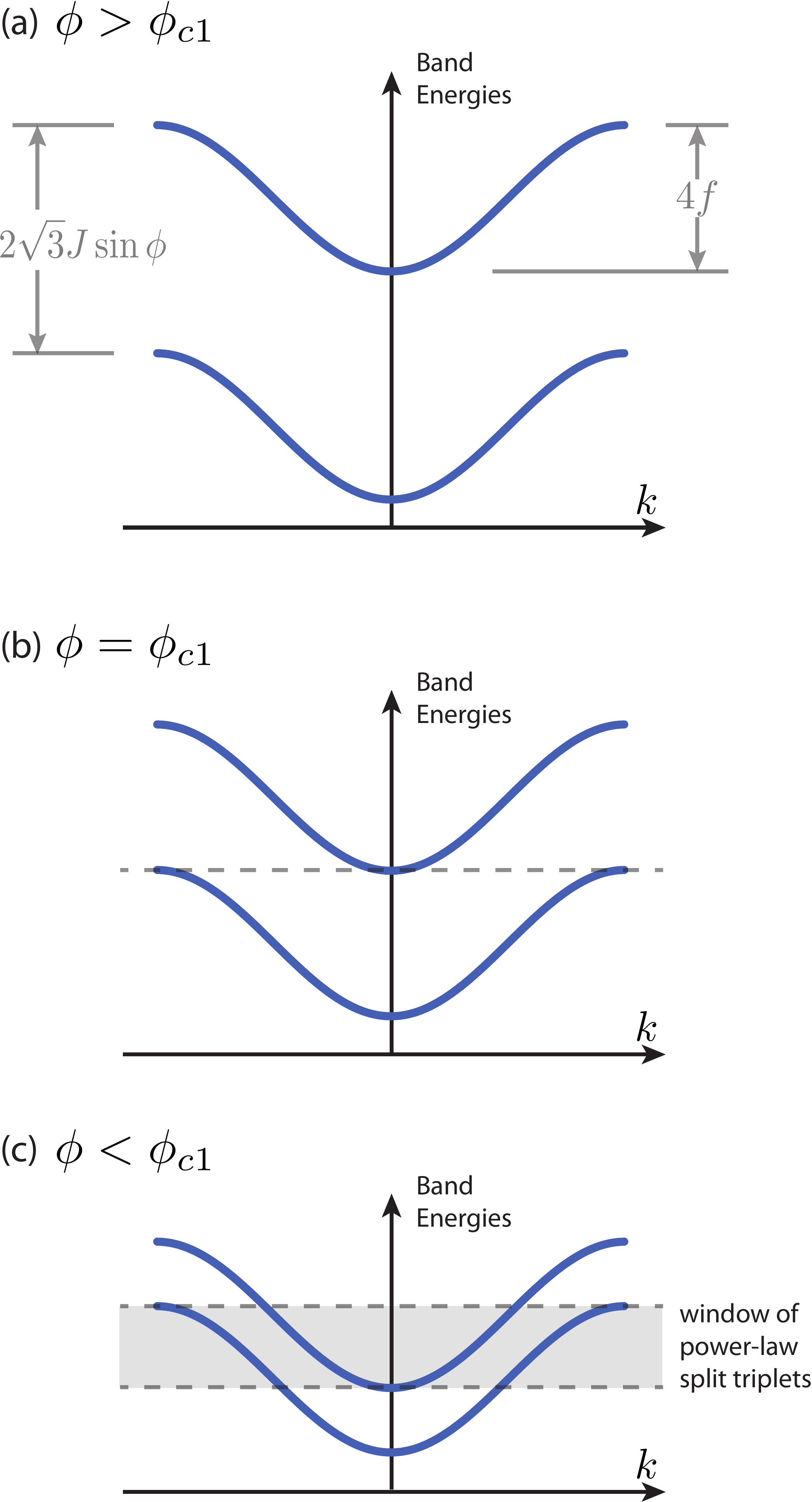}
\caption{Single-wall band energies for the chiral case near the ferromagnetic point; see Eq.~\eqref{bands}.  The upper band $E_+(k)$ arises from hopping of $\{|A|B\rangle, |B|C\rangle, |C|A\rangle\}$ states, while the lower band $E_-(k)$ similarly arises from $\{|B|A\rangle, |C|B\rangle, |A|C\rangle\}$ states.  (a) For $\phi$ greater than an $f$-dependent critical value $\phi_{c1}$ the bands are separated by a finite gap, leading to exponential splitting for all single-wall triplets.  (b) Indirect gapless excitations arise when $\phi = \phi_{c1}$.  Scattering from the system's edges mixes the resonant $k = 0$ and $k = \pi$ levels yielding power-law splitting among triplets at that energy.  (c) For $\phi> \phi_{c1}$ the bands overlap in a finite energy interval; all triplets within that window admit power-law splitting due to scattering at the edges.  }
\label{ChiralBands}
\end{figure}

Our aim is to now understand how these branches mix at the ends of a system with open boundaries.  
As Fig.~\ref{ChiralBands} illustrates, there are three distinct regimes distinguished by the degree to which the bands overlap.  
Implications for zero modes depend sensitively on the ratio of $f$ to the energy difference $E^+_{\rm wall}(\phi)-E^-_{\rm wall}(\phi) = 2\sqrt{3}J\sin\phi$.
It is thus often convenient to utilize the ratio
\begin{equation}
  \tilde \phi \equiv \sqrt{3}J\sin\phi/f 
  \label{phitilde}
\end{equation}
when analyzing the spectrum.

In the limit $\tilde \phi \gg 1$, the bands exhibit no overlap as shown in Fig.~\ref{ChiralBands}(a). The hybridization between the upper and lower branches at the edges of the chain thus can be treated perturbatively.  
If we revisit the process sketched in Fig.~\ref{fig:oneWallDiag1}, a clear qualitative difference from the ferromagnetic case appears.  
Now, when the $|A|B\rangle$ domain wall (originating from the upper branch) tunnels to the right end of the chain, conversion to $|A|C\rangle$ necessitates moving  to the energetically well-separated lower band.
Returning the $|A|C\rangle$ domain wall to the left end---where it can transition back to a $|B|C\rangle$ wall in the upper branch---requires the system to remain in the lower-energy space for a macroscopic number of steps.
The key physics is that chirality precludes resonantly switching domain-wall flavors at the system's boundaries.
Thus such processes produce exponential rather than power-law splitting among all triplets of single-wall eigenstates.
Numerical simulations of the projected Hamiltonian confirm this picture.  The yellow diamonds in Fig.~\ref{fig:oneWallED} illustrate the natural log of the splitting for the lowest three single-domain states versus system size assuming $\tilde{\phi} = 10$; the splitting is well-captured by an exponential.

The obvious indications for the absence of localized zero modes that we uncovered in the ferromagnetic limit thus disappear upon introducing a small chiral phase $\phi$ in the $\tilde \phi \gg 1$ regime.  
This finding is in harmony with the construction in Ref.\ \onlinecite{Fendley} of edge zero modes at $\phi\ne 0$ and with $f/J$ sufficiently small. 
Our results here do not prove, however, that zero modes persist to finite $f/J$, as sub-exponential splitting could arise from various other sources including tunneling within multi-domain-wall subspaces and higher-order processes in perturbation theory.  

Further observations do nevertheless suggest that localized zero modes indeed persist. First, we have studied numerically the effective Hamiltonian projected onto the two-domain-wall subspace and again found only exponential splitting when $\tilde \phi \gg 1$.
Second, it is possible to address higher-order processes rather efficiently by developing a set of Feynman-diagram-like rules governing domain wall dynamics.
Appendix \ref{Diagrams} describes the methodology.
There we briefly sketch a calculation indicating that through second order in $f/J$ the splitting amongst single-wall states remains exponential.
This is a result of the cancellation of contributions from various
diagrams related by symmetries of the allowed tunneling processes.
Though we have not proven that the cancellation occurs at all orders, the symmetries by which it works at low orders suggest that this is the case.

Lowering $\tilde \phi$ brings the upper and lower branches closer to one another in energy until they eventually begin to overlap at a critical $\tilde \phi_{c1}$.  Indirect gapless inter-band excitations are then permitted as shown in Fig.~\ref{ChiralBands}(b).  
The critical value follows from the condition that $E_+(k = 0,\phi) = E_-(k = \pi,\phi)$, yielding $\tilde \phi_{c1} = 2$.
This value corresponds to an $f$-dependent critical chiral phase 
\begin{equation}
  \phi_{c1} = \sin^{-1}\left(\frac{2f}{\sqrt{3}J}\right).
  \label{phic1}
\end{equation}
With open boundaries the chain's edges provide the momentum transfer needed here to \emph{resonantly} scatter $k = 0$ domain walls in the upper branch into $k = \pi$ walls in the lower branch.  
Processes like that of Fig.~\ref{fig:oneWallDiag1} consequently yield power-law splitting among the triplets at the bottom of the upper band and top of the lower band.  

With $\tilde \phi < \tilde\phi_{c1}$ the bands overlap in a finite energy window [see Fig.~\ref{ChiralBands}(c)] within which all triplets similarly admit power-law splitting.  
This picture is supported by our numerics in Fig.~\ref{fig:oneWallED} for $\tilde \phi = 0.01$, which reveal splitting that is exponential for the lowest excited triplet (squares) yet power-law for triplets at the middle of the lower branch (diamonds).  
In the perturbative regime, we can therefore conclusively rule out localized zero modes not just in the ferromagnetic limit, but in fact along the finite interval $0 \leq \phi \leq \phi_{c1}$.  

Mapping the problem to an `unfolded' system as described in the previous subsection and Appendix \ref{ProjectedH} provides an enlightening perspective.  
The projected single-wall Hamiltonian for the open chain again effectively describes a particle hopping in a periodic system of length $6(L-1)$.  Due to chirality, however, the unfolded Hamiltonian now includes a square-well potential that alternates between $+\sqrt{3}J \sin\phi$ and $-\sqrt{3}J \sin\phi$ every $L-1$ sites.  
For simplicity let us consider small chiralities and focus on states near the bottom of the upper and lower branches so that the particle's kinetic energy can be approximated with a free-particle dispersion $\propto k^2$.  
(Analogous arguments follow for states near the top of the bands.)  
Given sufficient kinetic energy to overcome the square-well barrier, the particle will efficiently sample the entire extended periodic chain.
Translating back to Potts model language, this means that domain walls can resonantly convert between different flavors at the edges of the system---resulting in power-law splitting at such energies.  
With insufficient kinetic energy the particle will remain predominantly within the square-well minima, decaying evanescently into the regions corresponding to square-well maxima.  
The evanescent decay translates into inefficient domain wall conversion and a splitting of triplets that decays exponentially with system size.

Suppose next that the chiral phase $\phi$ is slightly below $\pi/3$ so that the system resides close to the antiferromagnetic point.  
Because $E^-_{\rm wall}(\phi) \ll E^+_{\rm wall}(\phi)$ in this regime, the ordering of domain-wall subspaces changes dramatically compared to the cases examined above.  
At $f = 0$ the lowest-energy subspace consists of $|B|A\rangle, |C|B\rangle, |A|C\rangle$ states while the second-lowest subspace arises from the two-domain wall states $|C|B|A\rangle, |A|C|B\rangle, |B|A|C\rangle$. 
By contrast the states $|A|B\rangle$, $|A|C\rangle, |B|A\rangle$ that we often invoked earlier lie much higher up in the spectrum so that conversion processes sketched in Fig.~\ref{fig:oneWallDiag1} are now irrelevant.

We can, however, identity a new series of domain-wall processes that yield power-law splitting over a range of $\phi$ values. As an example, one can convert $|B|A\rangle$ to an $|A|C\rangle$ domain wall related by a global $\mathbb{Z}_3$ transformation by utilizing the lowest-lying two-wall subspace.  
This is accomplished by moving the original domain wall to the system's right end and then inserting a second domain wall, i.e., $|B|A\rangle \rightarrow |B|A|C\rangle$.   
Upon hopping both domain walls to the other end one can remove the leftmost wall, sending $|B|A|C\rangle \rightarrow |A|C\rangle$.
The remaining domain wall, having switched flavors, finally returns to its original location.
Such processes generically split the degeneracy among single-wall triplets in the lowest excited subspace.

To quantify the induced splitting it is once again useful to project onto the relevant subspaces. For periodic boundary conditions on the resulting effective Hamiltonian, the single-wall states broaden into a band with energies $E_-(k,\phi)$ given in Eq.~\eqref{bands}. The two-wall states are similar, although the energies for the latter are nontrivial since domain walls interact.
Introducing open boundaries enables transitions between these subspaces at the system's edges, which can supply any momentum necessary.  
Power-law splitting arises for triplets in the lower band \emph{if} they overlap in energy with the upper band.  
The critical $\phi_{c2}$ at which power-law behavior first appears follows from $E_-(k = \pi,\phi) \approx E_-(k = 0,\phi) + E^-_{\rm wall}(\phi)$, where the right side approximates the energy needed to insert an extra domain wall into a zero-momentum single-wall state. 
(Note that this estimate neglects domain wall interactions; this is reasonable since they are dilute.)
We then obtain 
\begin{equation}
  \phi_{c2} \approx \pi/3-\sin^{-1}\left(\frac{2 f}{\sqrt{3}J }\right).
  \label{phic2}
\end{equation}  
Interestingly, the $f$ dependence is remarkably similar to Eq.~\eqref{phic1} despite the rather different processes involved.
In the interval $\phi_{c2} \leq \phi < \pi/3$ at least one excited triplet exhibits power-law splitting in the perturbative regime, ruling out localized zero modes here too.  

\subsection{Scaling behavior}
\label{Scaling}

In the previous subsections we gained quite a bit of mileage by studying an effective Hamiltonian projected onto the single-domain-wall subspace.  
This truncated model is expected to capture the dominant contribution to the splitting amongst triplets of low-lying excited states in the limit $f, J\sin\phi \ll J$.  
Through analytical arguments corroborated by numerics we concluded that chirality induces a sharp change in the level spacing among triplets of excited states.  

More precisely, at $\phi = 0$ we showed that all single-domain-wall triplets exhibit power-law splitting with system size for arbitrary non-zero $f/J$.
Introducing small non-zero $\phi$ immediately restores exponential splitting for a range of the lowest- and highest-energy triplets but preserves power-law splitting for all intermediate single-wall states.  
Further increasing $\phi$ shrinks the window of power-law-split states until exponential splitting fully returns beyond a critical value $\phi_{c1}$ given in Eq.~\eqref{phic1}---presumably restoring exact localized zero modes.    
It is tempting to view this behavior as indicating chirality-tuned critical behavior in the excited-state spectrum, despite the fact that no singular behavior arises in the ground-state sector.  
We now provide evidence for such a scenario by showing that our numerical results exhibit data collapse consistent with a simple scaling ansatz.  

\begin{figure}
	\centering
	\subfigure[]{
	\includegraphics[width = \columnwidth]{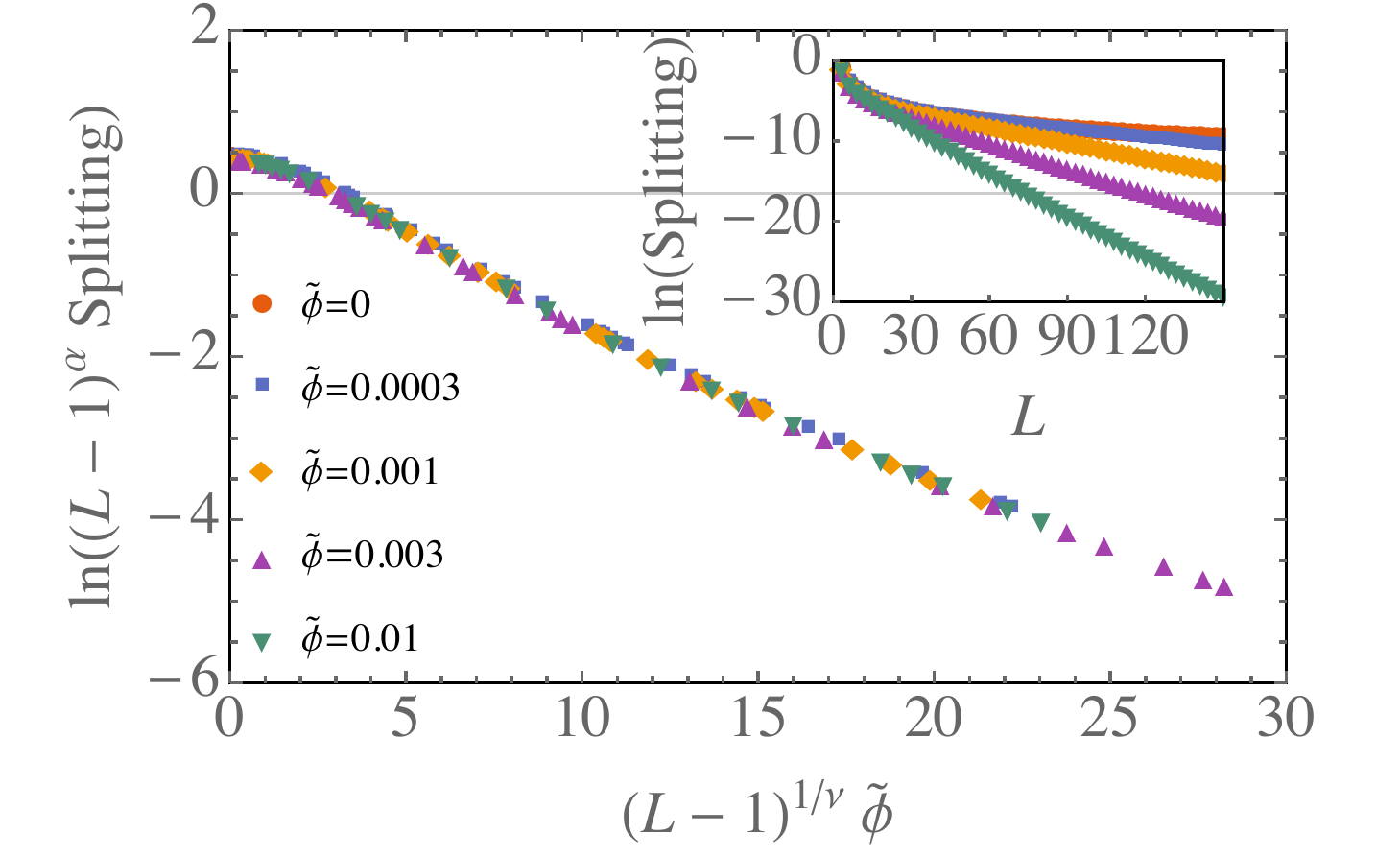}
	\label{fig:oneWallED3}}\qquad
	\subfigure[]{
	\includegraphics[width = \columnwidth]{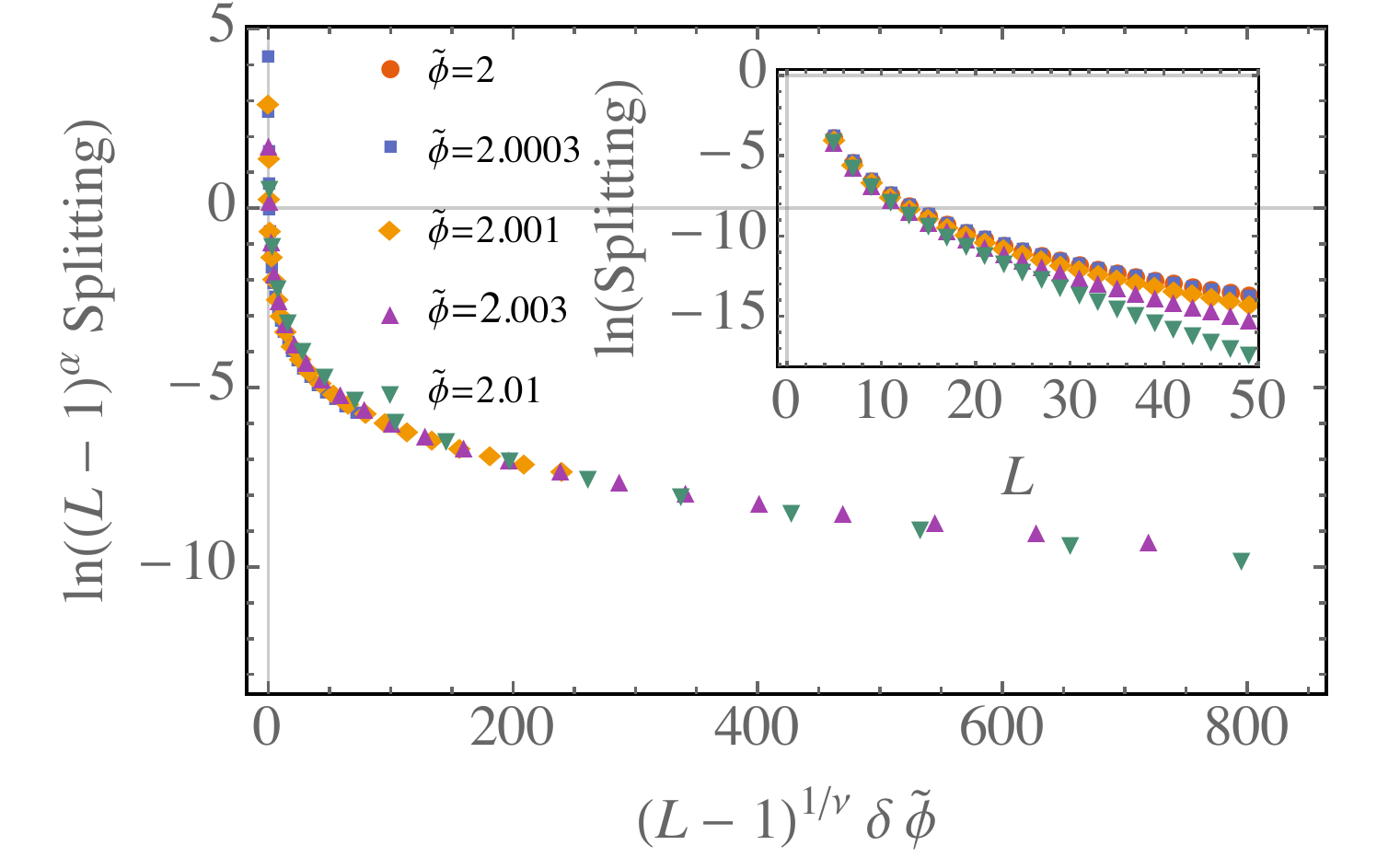}
	\label{fig:oneWallED2}}\qquad
	\caption{
	Demonstration of critical scaling in the excited-state spectra.  (a) Exact diagonalization results for the effective single-domain-wall Hamiltonian at various $\tilde\phi = \sqrt{3}J \sin\phi/f$ near the ferromagnetic limit. These data indicate chirality-driven critical behavior in the first excited triplet.  In accordance with Eq.~\eqref{Collapse} the data collapses onto a single curve when plotted versus $(L-1)^2\tilde\phi$ and upon rescaling the splitting by $(L-1)^2$, yielding the exponents $\alpha = 2$ and $\nu = 1/2$.
	The inset shows the natural log of the splitting among the lowest three excited states versus system size, illustrating the crossover between power-law and exponential splitting.
			  (b) Similar results as in (a) for a higher excited triplet and with chiral phases just beyond $\tilde\phi_{c1}=2$.  Critical exponents here are $\alpha=1$ and $\nu\approx 0.31$.
	}
\end{figure}

Provided $f, J\sin\phi \ll J$, the single-domain-wall splitting $\Delta E_\textrm{1-wall}$ of a given triplet should depend only on $\tilde \phi$ from Eq.~\eqref{phitilde} and the effective system size $\tilde L \equiv L-1$ for domain walls. Let us consider a particular triplet that transitions from exponential to power-law splitting at some (state-dependent) critical value $\tilde \phi_c$.  
We stress that $\tilde \phi_c$ is distinct from $\tilde \phi_{c1,2}$ discussed in the previous subsection; the former applies to just one particular multiplet.
Under rescaling of the length by $b$ we postulate the following scaling form,
\begin{equation}
  \Delta E_\textrm{1-wall}(b^{1/\nu} \delta \tilde \phi,\tilde L/b) = b^{\alpha} \Delta E_\textrm{1-wall}(\delta\tilde \phi,\tilde L)
\end{equation}
where $\delta \tilde \phi$ denotes the deviation from $\tilde \phi_c$ and $\nu,\alpha$ are critical exponents to be determined.  
Setting $b = \tilde L$ then allows us to write
\begin{equation}
  \Delta E_\textrm{1-wall}(\delta \tilde \phi,\tilde L) = \tilde L^{-\alpha}\mathcal{E}(\tilde L^{1/\nu} \delta\tilde \phi)
  \label{Collapse}
\end{equation}
for some function $\mathcal{E}$.  

As a concrete example consider the lowest-lying excited triplet, for which $\tilde \phi_c = 0$.  
Our results for the ferromagnetic case imply that in the perturbative regime $\alpha = 2$ while $\mathcal{E}(0)$ is some non-zero constant.\footnote{Of course in the extreme limit $f=0$ we have $\mathcal{E}(0)=0$.  Note also that the scaling forms provided here assume $f/J \ll 1$.  For larger transverse fields mixing between other sectors matters, and $\Delta E_\textrm{1-wall}$ hence becomes a function of three parameters---$\tilde \phi, L$, and $f/J$.  }
The remaining exponent $\nu$ follows from our simulations.  
In the inset of Fig.~\ref{fig:oneWallED3} we show the natural log of the splitting for the lowest single-wall triplet versus system size for a variety of $\tilde \phi$ values, illustrating the change from power-law to exponential behavior.
The same data appear in the main plot, but with the splitting scaled by $\tilde L^2$ and with the horizontal axis representing $\tilde L^2\tilde \phi$.  
Very clean data collapse is evident, confirming that our scaling ansatz holds with an exponent $\nu = 1/2$.
Figure~\ref{fig:oneWallED2} displays analogous scaling results for a triplet at the middle of the spectrum and with $\tilde\phi_c$ close to $\phi_{c1}$  [recall Fig.~\ref{ChiralBands}(b)].
Different exponents appear here, with $\alpha = 1$ and $\nu \approx 0.31$, but we again see evidence of critical behavior in the excited states.  

\subsection{Antiferromagnetic limit, \texorpdfstring{$\phi = \pi/3$}{phi = pi/3}}
\label{AFMpotts}

Finally, we consider the antiferromagnetic limit where $\phi = \pi/3$.  
One can anticipate that this case is rather special from the $f = 0 $ excitation energy quoted in Eq.~\eqref{Echiral}; in particular, with $\phi = \pi/3$ half of the domain-wall flavors cost no energy, resulting in a macroscopic ground-state degeneracy in the $f = 0$ limit.  
Here localized zero modes disappear for \emph{any} non-zero $f$ in the perturbative regime, just as in the ferromagnetic case (and for similar reasons).  Indeed, the construction of edge zero modes in Ref.\ \onlinecite{Fendley} works only for $f<J\sin(3\phi)$, and so fails at the antiferromagnetic point.

To demonstrate this result directly we need only examine the $f = 0$ ground-state manifold perturbed by infinitesimal $f$.  
It is convenient here to perform a gauge transformation $\sigma_x\rightarrow \omega^x \sigma_x$, which effectively transforms the Potts Hamiltonian to one with $\phi = \pi$.  
In the transformed model the $J$ term is minimized by any state for which no two adjacent spins align with one another (hence the `antiferromagnetic' nomenclature).
Thus the $f = 0$ limit supports $3\times2^{L-1}$ ground states.
Consider, for instance, a ground state of the form $|0101\cdots1010\rangle$. 
As usual we are interested in tunneling processes that connect states related by a global $\mathbb{Z}_3$ transformation, i.e., $|1212\cdots2121\rangle$, since such events split the degeneracy encoded by the exact zero modes present at $f = 0$.  
One possible pathway is for repeated action of the $\tau_x$ operators to first wind all of the even spins, sending $|0101\cdots1010\rangle \rightarrow |0202\cdots2020\rangle$, followed by all odd spins, sending $ |0202\cdots2020\rangle \rightarrow |1212\cdots2121\rangle$.
The system remains in the ground-state manifold throughout such processes, so that a splitting amongst ground-state triplets arises at first order in degenerate perturbation theory.

This splitting is thus expected to scale as a power-law with system size---similar to what we observed previously for the ferromagnetic case---precluding localized zero modes even for arbitrarily small non-zero $f$ as claimed.  
Interference effects that might alter this conclusion are absent since all non-zero matrix elements in the ground-state manifold are unity.  
Higher-order processes that require exiting the ground-state manifold are also negligible in the limit considered here.

There is a complementary way of arguing for power-law splitting in the antiferromagnetic case.
In the perturbative regime one can obtain an effective low-energy Hamiltonian for the chain by projecting the operator $\tau_i+\tau_i^\dagger$ appearing in the $f$ term onto the $f = 0$ ground-state manifold.  (The $J$ term of course projects trivially.)  
To do so it is useful to define bond operators that measure whether the spin at a given site is wound `clockwise' or `counterclockwise' relative to its neighbor.  
More specifically, we can introduce Pauli matrices $\eta^{z}_{i+1/2}$ such that $P \sigma_i^\dagger \sigma_{i+1}P = \omega^{\eta^z_{i+1/2}}$ with $P$ the ground-state projector.  
As an example $\eta^z_{i+1/2} = +1$ for the spin pair $|0_i 1_{i+1}\rangle$ while $\eta^z_{i+1/2} = -1$ for the mirrored configuration $|1_i 0_{i+1}\rangle$.
With this notation one can see that $\tau_i + \tau_i^\dagger$ flips both $\eta^z_{i-1/2}$ and $\eta^z_{i+1/2}$ if they have opposite signs, and otherwise projects trivially.  
More formally, we obtain
\begin{align}\begin{split}
	P(\tau_i + \tau_i^\dagger)P &= \frac{1}{2}\eta_{i-1/2}^x\eta_{i+1/2}^x\left(1-\eta_{i-1/2}^z\eta_{i+1/2}^z\right)
\\	&= \frac{1}{2}\left(\eta^x_{i-1/2}\eta^x_{i+1/2}+\eta^y_{i-1/2}\eta^y_{i+1/2}\right).
\end{split}\end{align}
Thus the bulk maps onto an XY model\footnote{We are grateful to David Clarke for pointing out to us this XY model mapping.}---which is gapless, making the power-law splitting argued above extremely natural.
The vanishing bulk excitation gap strongly suggests the generic absence of localized zero modes in the antiferromagnetic limit.

\section{Non-Perturbative Regime}
\label{sec:nonperturb}

Both the analytical arguments and numerical diagonalization exploited in Sec.~\ref{PerturbativeRegime} break down when $f/J$ becomes of order unity.  
Here we complement our earlier perturbative analysis using extensive DMRG simulations. 
To understand the nature of the spectrum in the non-perturbative regime, we computed the splitting among the ground states and first-excited triplet for a wide range of $\phi$ and $f/J < 1$ over which DMRG exhibits good convergence.  
The simulations turn out to perform particularly well at larger $f/J$.\cite{White-1992, Schollwoeck:Review05, McCulloch-2007}  
System sizes were taken to be at least $L = 10$ to minimize finite-size effects but sufficiently small to avoid the splitting falling below machine precision; the maximum $L$ considered varies from $19$ to $34$ depending on the Hamiltonian parameters. More details of our simulations are described in Appendix \ref{app:DMRG}.

\begin{figure*}
\centering
\subfigure[]{
\includegraphics[width=\columnwidth]{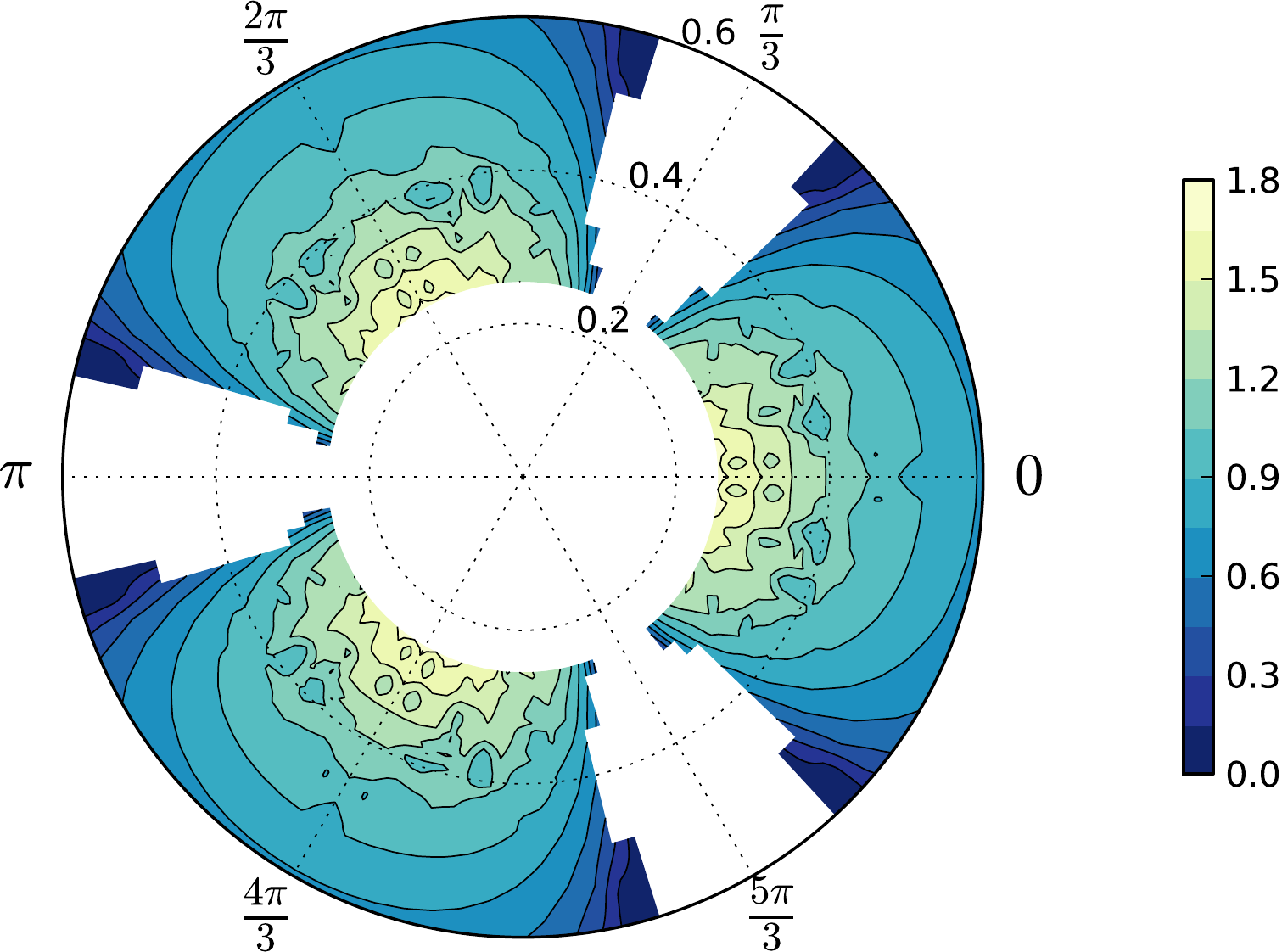}
\label{fig:phases1}}
\subfigure[]{
\includegraphics[width=\columnwidth]{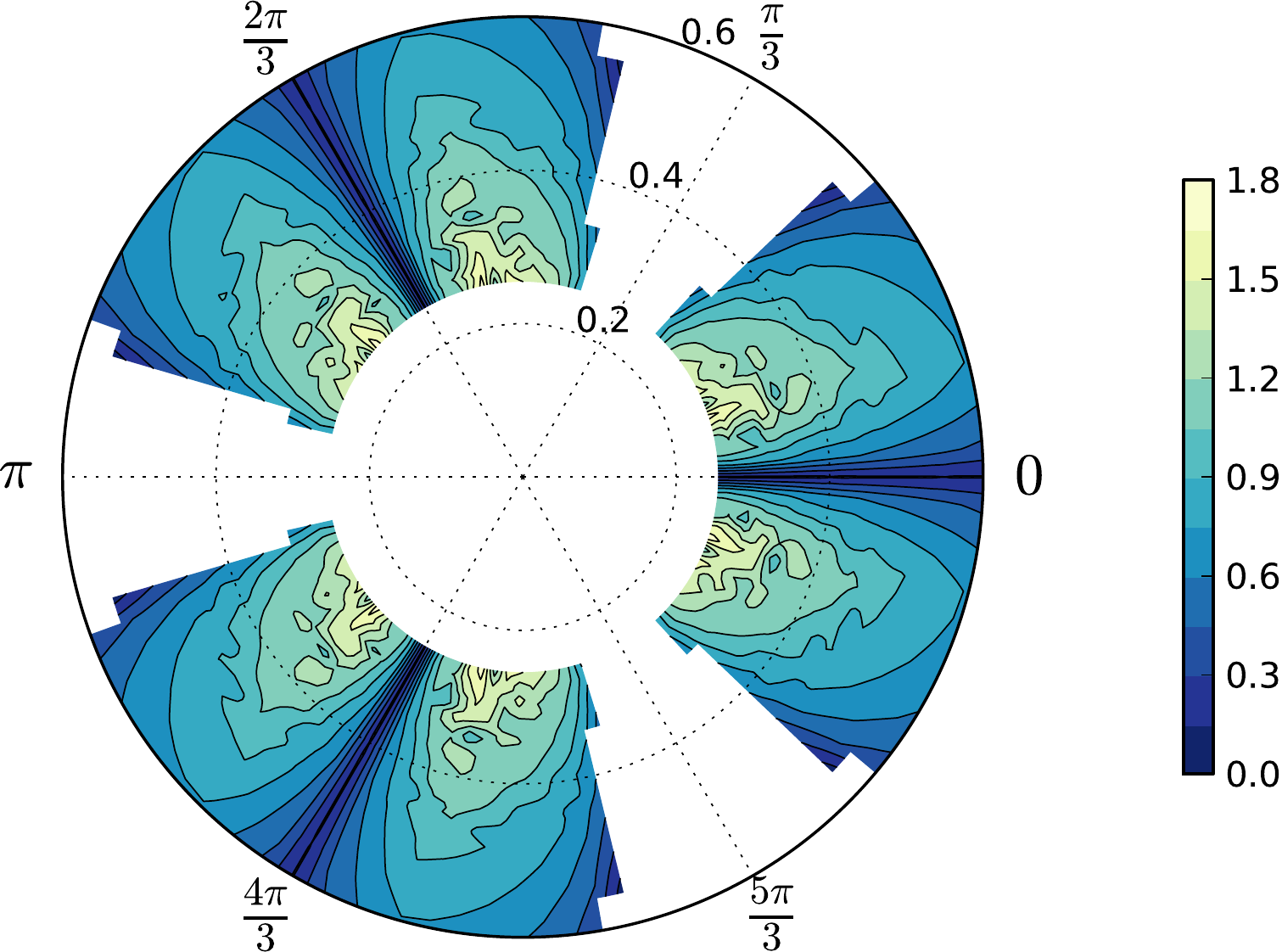}
\label{fig:phases2}}
\subfigure[]{
\includegraphics[width=\columnwidth]{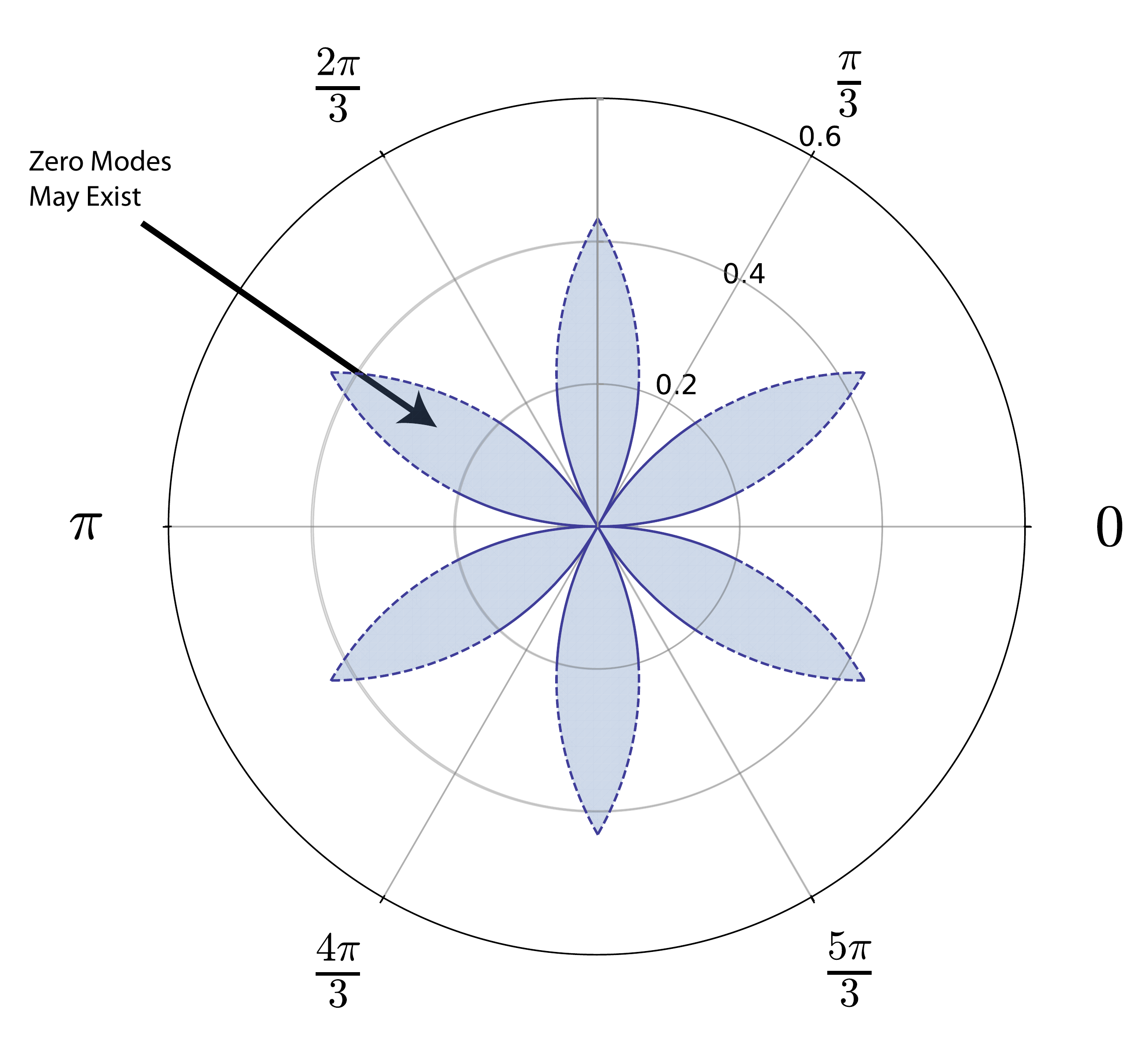}
\label{fig:phases3}}\qquad
\caption{(a) Ground-state and (b) lowest-lying excited-state splitting characteristics obtained from DMRG for various $f/J$ (radius) and $\phi$ (angle).  The splitting versus system size was fit to an exponential $e^{-\kappa L}$; the color represents the optimal $\kappa$ for each $(f/J, \phi)$ point.   For the ground states clear exponential splitting characterized by finite $\kappa$ appears throughout.  The excited states, in contrast, are characterized by $\kappa = 0$ in the ferromagnetic limit $\phi = 0$ (mod $2\pi/3$), suggesting power-law splitting as in the perturbative regime.  Omitted regions correspond to parameters where DMRG exhibited poor convergence. (c) Rough estimate for regions of $\phi$ and $f/J$ where zero modes are anticipated to exist based on extrapolation of our results for the perturbative regime.  The boundaries correspond to the critical values $\phi_{c1}$ and $\phi_{c2}$ respectively specified in Eqs.~\eqref{phic1} and \eqref{phic2}.  Dotted lines roughly indicate the non-perturbative regime where we expect quantitative corrections to the extrapolation.  Notably, the lobe structure mimics the regions where Fendley explicitly constructed localized zero-mode operators\cite{Fendley}.}
\label{fig:phases}
\end{figure*}

To quantify the splittings obtained from the DMRG, our data were fit to an exponential form $\propto e^{-\kappa L}$.
The color scales in Figs.~\ref{fig:phases1} and \ref{fig:phases2} show the optimal $\kappa$ value versus $f/J$ (radius) and $\phi$ (angle) for the ground-state and first excited splittings, respectively.
The ground states exhibit an exponential splitting characterized by a non-zero $\kappa$ throughout the regime of convergence, which notably does not include the antiferromagnetic limit because of its gaplessness.
As in the perturbative regime, however, richer physics arises for the first excited triplet.
In the ferromagnetic limit we obtain $\kappa = 0$, indeed consistent with the power-law splitting found in the perturbative regime.  
For the chiral cases, by contrast, $\kappa$ takes on finite values indicating exponential splitting with system size, also in agreement with our perturbative analysis.

We showed earlier that power-law splitting for excited states survives over a \emph{finite} range of chiral phases $\phi$.  
However, accessing these levels within DMRG rapidly becomes prohibitive as one increases the energy.
Thus the finite-$\phi$ transitions are difficult to capture numerically.  
Instead we crudely estimate the global zero-mode stability regions by naively extrapolating our results for the critical $\phi$ values obtained for the perturbative regime in Sec.~\ref{Chiralpotts}. 
Figure \ref{fig:phases3} displays the results of our extrapolation.
The shaded regions indicate the parameter space over which zero modes are expected to survive based on our perturbative criterion.  
While of course this scheme is not quantitatively reliable, we do expect to capture the qualitative trends.
[One feature that is \emph{not} expected to be robust is the `accidental' symmetry of Fig.~\ref{fig:phases3} with respect to sending $\phi \rightarrow \pi/6-\phi$, as is clear from the numerics in Fig.~\ref{fig:phases2}. This pathological property should be absent even in the perturbative regime upon including interactions between domain walls, which were neglected in our estimate of $\phi_{c2}$.]

It is interesting to discuss these findings in relation to Ref.~\onlinecite{Fendley}, where using an iterative method localized zero modes were constructed explicitly in the limit where the control parameter
\begin{equation}
	r \equiv \frac{f}{J\sin\left(3\phi\right)}
\end{equation}
is much less than one.  
In this case corrections to the $f = 0$ zero modes could be arranged in a series that clearly decays exponentially into the bulk of the chain.  
Remarkably, our extrapolation in Fig.~\ref{fig:phases3} follows the lobe-like form of the control parameter $r$---strongly suggesting that we have correctly identified the essential physics that determines the robustness of exact zero modes in parafermion chains.

\section{Conclusions}
\label{Conclusions}

In this paper we employed a variety of techniques to diagnose the puzzling stability of local parafermion zero modes---which differs markedly from the Majorana case---identified by Ref.~\onlinecite{Fendley}.  
Viewing the physics from the mathematically equivalent lens of the chiral three-state Potts model proved particularly illuminating.  
In Potts language physically intuitive domain-wall hopping/conversion processes produce power-law splitting among low-lying excited states in the ferromagnetic and antiferromagnetic limits of the model, ensuring the generic demise of localized parafermion zero modes in both cases.  
Chirally deforming the Hamiltonian tends to suppress these processes, and for sufficiently large deviations from the ferromagnetic/antiferromagnetic limit restores exponential splitting of all low-lying excited states that we examined.  
We speculate that the restoration of exponential splitting coincides with the emergence of localized zero modes; although a general proof remains unavailable such a scenario gels nicely with the results from Ref.~\onlinecite{Fendley}.

We also showed that the transition from power-law to exponential behavior reflects a subtle type of chirality-tuned quantum criticality that, interestingly, emerges only at energies above the ground states.  
Indeed, except in the antiferromagnetic limit the ground states remain degenerate up to exponentially small corrections in all cases and do not exhibit any singular behavior as a function of chirality.  

Although we focused on $\mathbb{Z}_3$ parafermion chains for simplicity, many of our results extend straightforwardly to $\mathbb{Z}_{N>3}$ systems (where similar stability issues arise\cite{Fendley}).  
For instance, it is clear that here too domain-wall tunneling events generically preclude exponentially localized modes in the ferromagnetic and antiferromagnetic limits.  
Our findings also naturally explain the comparative robustness of localized Majorana zero modes in the Kitaev chain, since the relevant domain-wall processes that we invoked for the Potts model have absolutely no analogue in that context.  

Since we motivated this work from the vantage point of topological quantum computation, it is worth re-emphasizing that the existence of localized zero modes as defined here is most certainly overkill for this application.  
Harnessing non-Abelian statistics in parafermion chains (and related systems) merely requires degenerate \emph{ground states} of a topological phase to within exponential accuracy; this weaker condition appears much more broadly as noted above, even when exact zero mode operators are definitely absent.  
Systems supporting exact zero modes do, nevertheless provide the appealing possibility of performing topological quantum computation \emph{at finite energy density}---a possibility first proposed in the framework of many-body localization\cite{HuseMBL} (see also Ref.~\onlinecite{Bahri}).
We suggest that parafermion chains offer particularly interesting platforms to explore in this regard.
Apart from chirality, localization via quenched disorder should provide another means of suppressing the domain-wall processes that produced power-law splitting in the ferromagnetic limit.  
It would be quite interesting to perform large-scale exact-diagonalization studies to explore this scenario further in future work.

\acknowledgments{
We are indebted to David Clarke, Nate Lindner and Lesik Motrunich for illuminating conversations, as well as to Miles Stoudenmire for invaluable discussions on numerics.
We also acknowledge funding from the NSF through grants DMR-1341822 (A.~J.\ \& J.~A.) and DMR/MPS1006549 (P.~F.); the Alfred P.\ Sloan Foundation (J.~A.); the Sherman Fairchild Foundation (R.~M.); the Caltech Institute for Quantum Information and Matter, an NSF Physics Frontiers Center with support of the Gordon and Betty Moore Foundation; the Caltech Summer Undergraduate Research Fellowship program along with partial support from the family of Jean J.\ Dixon (A.~J.); and the Walter Burke Institute for Theoretical Physics at Caltech.
}

\appendix
\section{Path cancellations at higher orders in perturbation theory}
\label{Diagrams}

Section \ref{PerturbativeRegime} performed a perturbative analysis that incorporated the lowest-order processes that produced a splitting among eigenstates with different trialities.  
For excited states it is possible to capture very similar transformations to that of Fig.~\ref{fig:oneWallDiag1} in far fewer perturbation steps if one considers terms higher order in $f/J$.  
One thus might worry that such processes, while parametrically smaller in $f/J$, nevertheless produce the dominant scaling with system size.
In the chiral case with $\phi_{c1}<\phi<\phi_{c2}$, for instance, if power-law splitting were to appear at higher orders then that would immediately imply a further reduction in stability window for localized zero modes.  
However, our DMRG results---which are non-perturbative and hence include \emph{all} orders---suggest that this is not the case since our simulations captured exponential splitting among the lowest excited triplet for any non-zero chirality.  
In this appendix we provide more support for this conclusion by examining higher-order perturbative computations.

While cumbersome, higher-order calculations can be greatly facilitated by using of a diagrammatic representation of domain-wall events similar to Fig.~\ref{fig:oneWallDiag1}.  
For this purpose it is useful to attach an upward-pointing arrow to lines indicating domain walls of the type $|A|B\rangle$, $|B|C\rangle$, and $|C|A\rangle$ and a downward arrow for the other three flavors.  
The following vertex rules then arise:
\begin{enumerate}
\item
The boundary of the system acts as a source and sink for domain walls. This is the one-line vertex.
\item
Two lines may be created or destroyed at a point if and only if they have opposite arrow directions. This is the two-line vertex.
\item
A three-line vertex may exist if and only if all of the arrows are incoming or outgoing.
\item
In a given perturbation step each line may either remain fixed or move one spatial unit.
\end{enumerate}
Since all matrix elements in the perturbation theory are either 0 or 1 the assignment of weights to diagrams conforming to these rules is straightforward.  
Any step that moves the system out of the original subspace gets penalized by a factor of $\pm f/|\delta E|$ where $|\delta E|$ is the magnitude of the energy change incurred.  
The $+$ sign arises if the system moves to a higher-energy state, while the $-$ sign arises if a lower-energy state results.

\begin{figure}
\centering
\subfigure[]{
\includegraphics[width=0.20\textwidth]{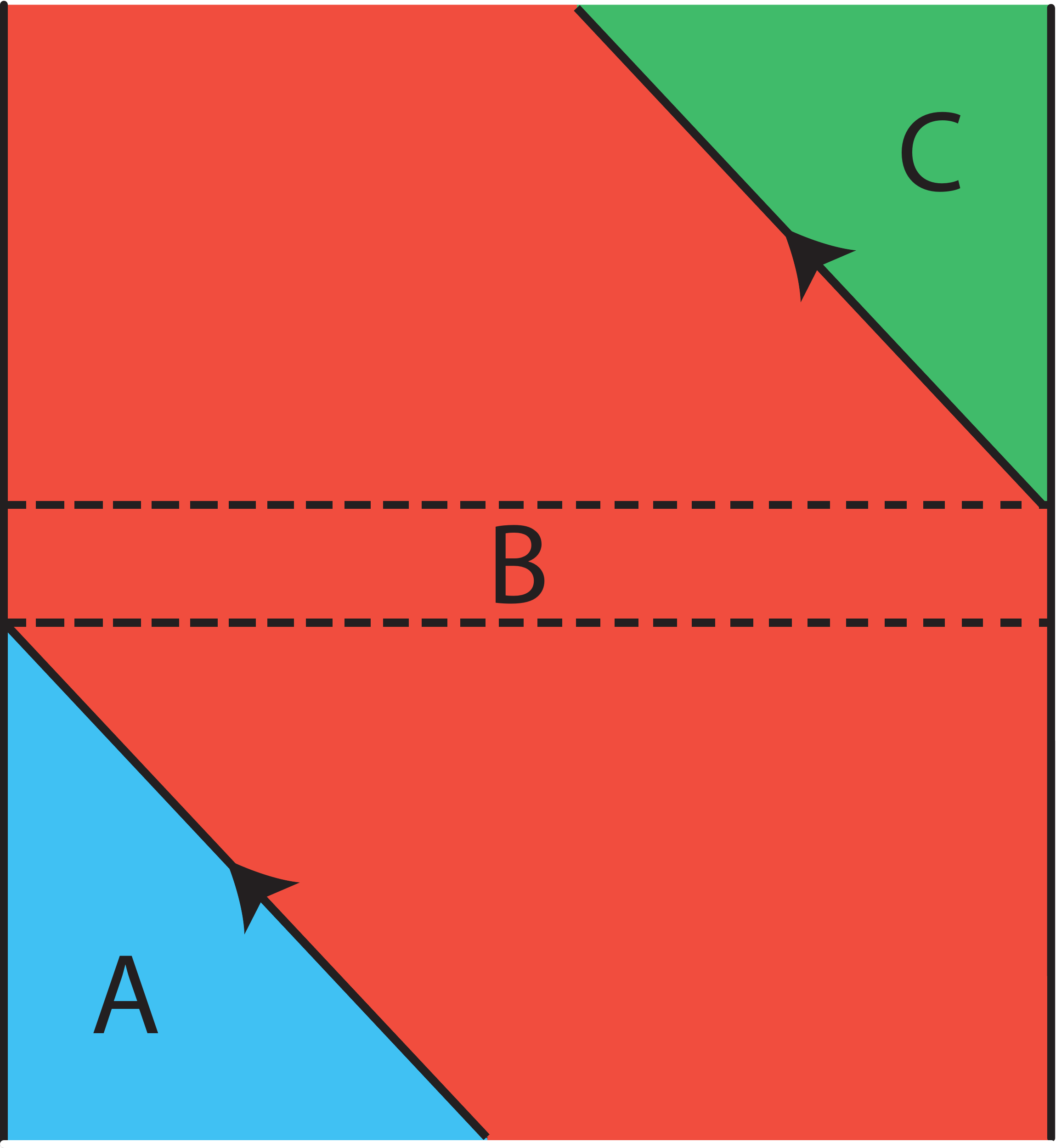}
\label{fig:cancel1}}\qquad
\subfigure[]{
\includegraphics[width=0.20\textwidth]{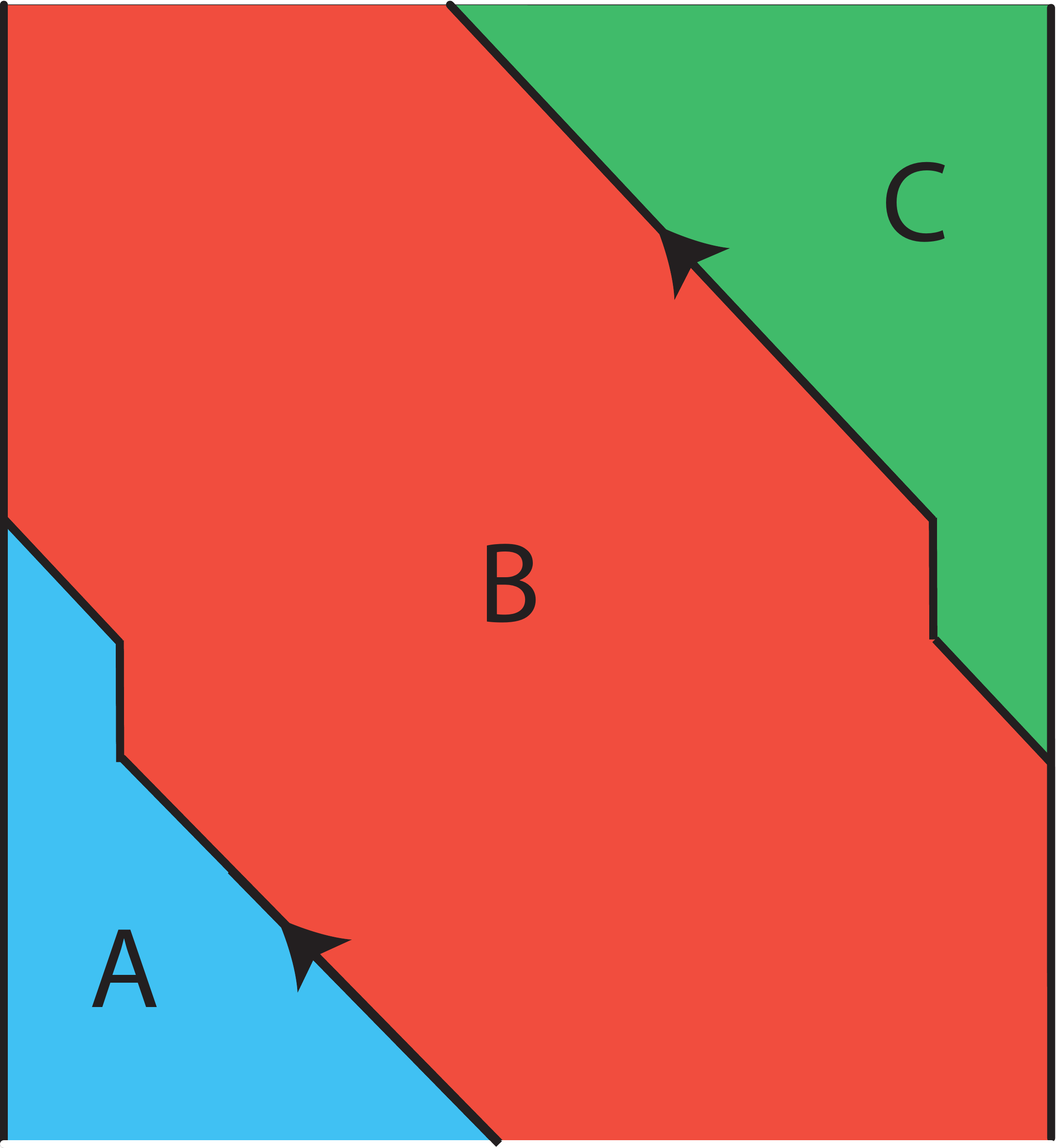}
\label{fig:cancel2}}\qquad
\caption{Diagrammatic representation of select higher-order processes that produce a splitting of excited-state triplets. Contrary to Fig. 4(b) these processes require leaving the single-domain-wall subspace.  Individually, the events (a) $|A|B\rangle \rightarrow |B\rangle \rightarrow |B|C\rangle$ and (b) $|A|B\rangle \rightarrow |A|B|C\rangle \rightarrow |B|C\rangle$ produce power-law splitting even in the chiral case, but their contributions cancel since they come with opposite signs.  Similar cancellations appear also at higher orders.  }
\end{figure}

Consider first the diagram in Fig.~\ref{fig:cancel1} depicting a second-order process wherein $|A|B\rangle \rightarrow |B\rangle \rightarrow |B|C\rangle$.  
By themselves, events like this yield power-law splitting $-c f^2/[L^2 E^+_{\textrm{1-wall}}(\phi)]$ with $c > 0$, \emph{even in the chiral case}.  
There is, however, a compensating process shown in Fig.~\ref{fig:cancel2} where $|A|B\rangle \rightarrow |A|B|C\rangle \rightarrow |B|C\rangle$.
Since here the system enters a higher-energy sector these events produce the exact opposite contribution $+c f^2/[L^2 E^+_{\textrm{1-wall}}(\phi)]$.
The existence of such cancellations is the main message of this appendix.
Preliminary calculations at third order point to a similar outcome, and
we expect that they are a generic feature of higher-order perturbation theory in certain regions of parameter space.
Again, this conclusion is supported by our DMRG results at intermediate $f/J$.  

\begin{widetext}
\section{Projection domain-wall Hamiltonians}
\label{ProjectedH}

In the limit $f,J\sin\phi \ll J$ one can to a good approximation neglect tunneling processes that mix sectors with different numbers of domain walls.  
Formally this is achieved by projecting $H$ in Eq.~\eqref{Hpotts} onto a subspace with fixed domain wall number.  The procedure is generally straightforward though some care is necessary at the boundaries.  
In the one-wall sector the resulting effective Hamiltonian admits a particularly clean block form:
\begin{align}
	H_{\textrm{1-wall}} =\begin{pmatrix}
		(V-Q) & -f I\\
		-f I & -Q & -f I\\
		 & -f I & -Q & \ddots\\
		 &  & \ddots & \ddots & -f I\\
		 &  &  & -f I & -Q & -f I\\
		 &  &  &  & -f I & (W-Q)
	\end{pmatrix},
\end{align}
with $I$ the $6\times6$ identity matrix,
\begin{align}
	V= -f \begin{pmatrix}
		0 & 0 & 0 & 0 & 0 & 1\\
		0 & 0 & 1 & 0 & 0 & 0\\
		0 & 1 & 0 & 0 & 0 & 0\\
		0 & 0 & 0 & 0 & 1 & 0\\
		0 & 0 & 0 & 1 & 0 & 0\\
		1 & 0 & 0 & 0 & 0 & 0
	\end{pmatrix} , \quad
	W= -f \begin{pmatrix}
		0 & 1 & 0 & 0 & 0 & 0\\
		1 & 0 & 0 & 0 & 0 & 0\\
		0 & 0 & 0 & 1 & 0 & 0\\
		0 & 0 & 1 & 0 & 0 & 0\\
		0 & 0 & 0 & 0 & 0 & 1\\
		0 & 0 & 0 & 0 & 1 & 0
	\end{pmatrix} , \quad\textrm{and}\quad
	Q=\sqrt{3}J\sin\phi\begin{pmatrix}
		1 & 0 & 0 & 0 & 0 & 0\\
		0 & -1 & 0 & 0 & 0 & 0\\
		0 & 0 & 1 & 0 & 0 & 0\\
		0 & 0 & 0 & -1 & 0 & 0\\
		0 & 0 & 0 & 0 & 1 & 0\\
		0 & 0 & 0 & 0 & 0 & -1
	\end{pmatrix}.
\end{align}
\end{widetext}

As noted in Sec.~\ref{FMpotts}, because of the form of the boundary terms described by $V$ and $W$ it is possible to recast $H_{\textrm{1-wall}}$ in terms an enlarged periodic chain of size $6(L-1)$.  
In this mapping the domain wall becomes a particle living on sites labeled by position $x$, with the domain-wall flavor corresponding to
$j \equiv {\rm ceiling}\left(\frac{x}{L-1}\right)$.  
We will take $j = 1,3,5$ to represent $|A|B\rangle$, $|B|C\rangle$, and $|C|A\rangle$, and $j = 2,4,6$ to represent $|A|C\rangle$, $|B|A\rangle$, and $|C|B\rangle$, respectively.
Let us now specialize to the ferromagnetic limit, $\phi = 0$.  
Denoting the position modulo $L-1$ by $y$, it is straightforward to verify that the eigenstates of the projected Hamiltonian are given by
\begin{equation}
|\psi_k\rangle = \sum_{j=1}^6 \sum_{y=1}^{L-1} e^{ik\left((L-1)j+y\right)}|j,y\rangle,
\label{eq:eigs}
\end{equation}
with eigenvalues
\begin{equation}
E(k)=-2f\cos(k).
\label{Ek2}
\end{equation}
These energies are the same as those in Eq.~\eqref{Ek} up to a constant; however, because of the enlarged system size in this description the allowed momenta are now $k\in \frac{2\pi}{6(L-1)}\mathbb{Z}$.
Note that the wavefunctions specified in Eq.~\eqref{eq:eigs} are also triality eigenstates with eigenvalue
\begin{equation}
Q = e^{2ik(L-1)}.
\end{equation}
[The action of $\hat{Q}$ simply maps $j\rightarrow(j+2) \bmod 6$.]

For extracting the splittings of interest, it is convenient to label the energies and momenta according to both the triplet $a$ to which they belong and their triality $Q$, i.e., as $E_{a,Q}$ and $k_{a,Q}$.
We need only specify that $k_{a,Q=\omega}=-k_{a,Q=\omega^2}>0$ and require that $\left|k_{a,Q=\omega}-k_{a,Q=1}\right|=\frac{\pi}{3(L-1)}$.
Using Eqs.~\eqref{Splitting} and \eqref{Ek2}, we then find
\begin{eqnarray}
  {\rm Splitting}[a] = 2\sqrt{2}f\left|\cos(k_{a,Q=1})-\cos(k_{a,Q=\omega})\right|,
\end{eqnarray}
which in the large-$L$ limit converges to Eq.~\eqref{eq:splittEq}.

For our numerical simulations of the truncated single-wall model, we explicitly split the spectrum into the three triality sectors by simultaneously diagonalizing the triality operator and the effective Hamiltonian.
In practice this may be done by adding the two operators together with random coefficients, determining the diagonalizing unitary operator for the combination, and then using it to diagonalize the two operators individually.
For almost all choices of random numbers this diagonalizes both operators.
Pathological cases may be handled separately, or by re-running the calculation.

To handle the two-wall sector, a Python+Cython+NumPy code was written to explicitly enumerate all states of interest and compute the relevant matrix elements of $H$ between them.
In either sector the effective Hamiltonians can be diagonalized numerically via standard routines to obtain splittings for any triplet of single-domain-wall states.

\section{DMRG methods}
\label{app:DMRG}

All DMRG computations in this paper utilized the Developer Branch of ITensor (\url{http://itensor.org}) commit 475352f76c6209db865ea4405cb86f665f40fae5.
A control file, a Hamiltonian file, and a model file were created based on existing ITensor code and with the assistance of Miles Stoudenmire, one of the authors of ITensor. These files extended ITensor to perform calculations on the $\mathbb{Z}_3$ Potts model, which is not a native function of the code. The Eigen C++ library version 3.0 was also used in the main control file for diagonalizing the states that ITensor produced. This generates excited states with less ground-state overlap than DMRG alone gives.

The results from ITensor were verified for small system sizes with the open-source Quantum Chains package (\url{https://github.com/adamjermyn/QuantumChains/}) written by the first author.  Any of the code used in this paper is available upon request from the first author.

\hbadness 10000
\bibliography{zeroModes}

\end{document}